\documentclass[aps, prd, floatfix, twocolumn, superscriptaddress, nofootinbib]{revtex4-1}

\usepackage{latexsym}
\usepackage{amsmath}
\usepackage{amssymb}
\usepackage{amsfonts}

\usepackage{dsfont}
\usepackage{slashed}
\usepackage{bm}
\usepackage{urwchancal}

\usepackage[vcentermath]{youngtab}

\usepackage{color}
\definecolor{purple}{rgb}{0.5,0,0.5}
\definecolor{blue}{rgb}{0.0,0,0.9}
\definecolor{prdblue}{rgb}{0.133,0.118,0.498}

\usepackage{enumitem}
\usepackage{multirow}
\usepackage{subfigure}
\usepackage{textcomp}

\usepackage{supertabular} 
\usepackage{placeins}
\usepackage{epsfig}
\usepackage{graphicx}

\begin{document}


\title{The structure of the X(3872) as explained by a Diffusion Monte Carlo calculation}

\author{M.C. Gordillo}
\email[]{cgorbar@upo.es}
\affiliation{Departamento de Sistemas F\'isicos, Qu\'imicos y Naturales, Universidad Pablo de Olavide, E-41013 Sevilla, Spain}

\author{F. De Soto}
\email[]{fcsotbor@upo.es}
\affiliation{Departamento de Sistemas F\'isicos, Qu\'imicos y Naturales, Universidad Pablo de Olavide, E-41013 Sevilla, Spain}

\author{J. Segovia}
\email[]{jsegovia@upo.es}
\affiliation{Departamento de Sistemas F\'isicos, Qu\'imicos y Naturales, Universidad Pablo de Olavide, E-41013 Sevilla, Spain}

\date{\today}

\begin{abstract}
Two decades after its unexpected discovery, the properties of the $X(3872)$ exotic resonance are still under intense scrutiny. In particular, there are doubts about its nature as an ensemble of mesons or having any other internal structure.
We use a Diffusion Monte Carlo method to solve the many-body Schr\"odinger equation that describes this state as a $c \bar c n \bar n$ ($n=u$ or $d$ quark) system. This approach accounts for multi-particle correlations in physical observables avoiding  the usual quark-clustering assumed in other theoretical techniques.
The most general and accepted pairwise Coulomb$\,+\,$linear-confining$\,+\,$hyperfine spin-spin interaction, with parameters obtained by a simultaneous fit of around 100 masses of mesons and baryons, is used. The $X(3872)$ contains light quarks whose masses are given by the mechanism responsible of the dynamical breaking of chiral symmetry. The same mechanisms gives rise to Goldstone-boson exchange interactions between quarks that have been fixed in the last 10-20 years reproducing hadron, hadron-hadron and multiquark phenomenology.
It appears that a meson-meson molecular configuration is preferred but, contrary to the usual assumption of $D^0\bar{D}^{\ast0}$ molecule for the $X(3872)$, our formalism produces $\omega J/\psi$ and $\rho J/\psi$ clusters as the most stable ones, which could explain in a natural way all the observed features of the $X(3872)$.
\end{abstract}



\maketitle


\noindent\emph{Introduction}.\,---\, A very successful classification scheme for hadrons in terms of their valence quarks and antiquarks was independently proposed by Murray Gell-Mann~\cite{GellMann:1964nj} and George Zweig~\cite{Zweig:1964CERN} in 1964. This classification was called the quark model, and it basically separates hadrons in two big families: mesons (quark-antiquark) and baryons (three-quark). The quark model received experimental verification beginning in the late 1960s and, despite extensive experimental searches, no unambiguous candidates for exotic configurations were identified until the turn of this century, with the discovery by the Belle Collaboration in 2003~\cite{Choi:2003ue} of the $X(3872)$ (also known as $\chi_{c1}(3872)$) in the invariant mass spectrum of $\pi^+ \pi^- J/\psi$ produced in $B^\pm \rightarrow K^\pm X(3872) \rightarrow K^\pm (\pi^+ \pi^- J/\psi)$ decays. Since then, more than two dozens of unconventional charmonium- and bottomonium-like states, the so-called XYZ mesons, have been observed at B-factories (BaBar, Belle and CLEO), $\tau$-charm facilities (CLEO-c and BESIII) and also proton-(anti)proton colliders (CDF, D0, LHCb, ATLAS and CMS). For an extensive presentation of the status of heavy quarkonium physics, the reader is referred to several reviews~\cite{Lebed:2016hpi, Ali:2017jda, Guo:2017jvc, Olsen:2017bmm, Liu:2019zoy, Brambilla:2019esw}.

Even today, the $X(3872)$ represents a puzzle with no consensus about its structure. Its current world average mass is $(3871.69\pm0.17)\,\text{MeV}$~\cite{Zyla:2020zbs}, very similar to that of the charmed meson pair: $m(D^0\bar{D}^{\ast0})=(3871.69 \pm 0.07)\,\text{MeV}$. This state seems also  extremely narrow with a width less than $1.2\,\text{MeV}$ at $90\%$ confidence level~\cite{Zyla:2020zbs}. Experimental analysis from Belle and CDF collaborations, which combine angular information and kinematic properties of the $\pi^+\pi^-$ pair, strongly favor the quantum numbers $J^{PC}=1^{++}$~\cite{Abe:2005ix, Abe:2005iya, Abulencia:2005zc, Abulencia:2006ma}. 

From constituent quark models~\cite{Ebert:2002pp, Barnes:2005pb, Segovia:2013wma}, the predicted masses of the $J^{PC}=1^{++}$ $c\bar c$ low-lying states do not fit the one of the $X(3872)$. Nonetheless, the strongest evidence against a $c\bar c$ assignment for the $X(3872)$ state is the fact that the di-pion mass distribution in the $X(3872)\to \pi^+ \pi^- J/\psi$ process proceeds through the $X(3872)$ decaying into a $\rho^0 J/\psi$ final state~\cite{Choi:2003ue, Abulencia:2005zc}, which would violate isospin conservation if the $X(3872)$ were interpreted as a conventional charmonium state.

The interpretation of $X(3872)$ as a molecular bound state with a very small binding energy~\cite{Tornqvist:1993ng, Close:2003sg, Swanson:2003tb, Voloshin:2003nt, Wong:2003xk, Fleming:2007rp, Guo:2017jvc} is favored by the closeness of the $X(3872)$'s mass and the $D^0\bar D^{\ast0}$ threshold. The ratio~\cite{Zyla:2020zbs} 
\begin{equation}
{\cal R}_{\omega-\rho} \equiv \frac{{\cal B}(X(3872)\to \pi^+\pi^-\pi^0 J/\psi)}{{\cal B}(X(3872)\to \pi^+\pi^- J/\psi)} = 1.1 \pm 0.4\,,
\label{eq:Romega}
\end{equation}
measured by Belle~\cite{Abe:2005ix}, BaBar~\cite{delAmoSanchez:2010jr} and BESIII~\cite{Ablikim:2019zio}, despite being well understood if the $X(3872)$ is interpreted as a $DD^{\ast}$ molecular state, is not well reproduced in theoretical calculations, \emph{e.g.}, ${\cal R}_{\omega-\rho}\approx0.15$ in Ref.~\cite{Matheus:2009vq}. This, together with the observation of $X(3872)$  decaying electromagnetically into $\gamma J/\psi$ and $\gamma \psi(2S)$ final states, could be interpreted as an indication that there is, at least, a significant mixing of the $c\bar c$ component with the $D^0 \bar{D}^{\ast0}$ molecule; actually, there are theoretical works exploring such possibility in the market~\cite{Eichten:2004uh, Suzuki:2005ha, Dong:2008gb, Ortega:2009hj}.

At this point, it seems clear that $X(3872)$ is not a charmonium state. Its minimal content must be then $4$ quarks, \emph{i.e.} a tetraquark system, an interpretation that was first proposed by Maiani \emph{et al.}~\cite{Maiani:2004vq}. Under that prism, the $X(3872)$ could appear as a bound state of a diquark-antidiquark cluster. This was based on the idea that diquarks can be treated as a confined quasi-particles and used as degrees-of-freedom in parallel with quarks themselves~\cite{Anselmino:1992vg, Close:2004ip, Selem:2006nd, Friedmann:2009mx, Barabanov:2020jvn}. However,  the drawback of the tetraquark picture is the proliferation of the predicted states~\cite{Maiani:2004vq} and the lack of selection rules that could explain why many of these states are not seen~\cite{Drenska:2008gr, Drenska:2009cd}.

In this letter we aim to elucidate the nature of the $X(3872)$ considering it as a tetraquark, specifically a $c\bar c n\bar n$ system, with $n$ labelling either $u$- or $d$-quark. Unlike other former studies, we are not going to assume any particular clustering between the valence quarks (antiquarks). Moreover, the interaction between them is the most simple and accepted one: Coulomb$\,+\,$linear-confining$\,+\,$hyperfine spin-spin, supplemented by general expressions of pseudo-Goldstone exchange interactions between light quarks due to dynamical breaking of chiral symmetry~\cite{Semay:1994ht, SilvestreBrac:1996bg}. Regardless of the breaking mechanism, the simplest Lagrangian which describes this situation must contain chiral fields to compensate the mass term and can be expressed as~\cite{Diakonov:2002fq}
\begin{equation}
{\cal L} = \bar{\psi}(i\, {\slash\!\!\! \partial} - M(q^{2}) U^{\gamma_{5}})\,\psi  \,,
\label{eq:ChiralL}
\end{equation}
where $U^{\gamma_{5}}=\exp(i\pi^{a}\lambda^{a}\gamma_{5}/f_{\pi})$, $\pi^{a}$ 
denotes nine pseudoscalar fields $(\eta_{0},\,\vec{\pi },\,K_{i},\,\eta _{8})$ 
with $i=1,\ldots,4$ and $M(q^2)$ is the constituent quark mass. The matrix of Goldstone-boson fields can be expanded in the following form
\begin{equation}
U^{\gamma _{5}} = 1 + \frac{i}{f_{\pi }} \gamma^{5} \lambda^{a} 
\pi^{a} - \frac{1}{2f_{\pi }^{2}} \pi^{a} \pi^{a} + \ldots
\label{eq:Ugamma5}
\end{equation}
The first term of the expansion generates the constituent quark mass while the
second gives rise to a one-boson exchange interaction between quarks. The
main contribution of the third term comes from the two-pion exchange interaction which has been simulated by means of a scalar exchange potential.

The expressions for the Goldstone-boson exchange potentials can be found in, for instance, Ref.~\cite{Segovia:2013wma}. Their parameters have been fixed in advance reproducing hadron~\cite{Valcarce:1995dm, Vijande:2004he, Segovia:2008zza, Segovia:2008zz, Segovia:2009zz, Segovia:2011zza, Segovia:2015dia, Ortega:2016mms, Segovia:2016xqb, Yang:2017qan}, hadron-hadron ~\cite{Fernandez:1993hx, Valcarce:1994nr, Ortega:2009hj, Ortega:2016hde, Ortega:2016pgg, Ortega:2016pgg, Ortega:2017qmg, Ortega:2018cnm, Ortega:2020uvc} and multiquark~\cite{Vijande:2006jf, Yang:2015bmv, Yang:2018oqd, Yang:2019itm, Yang:2020twg, Yang:2020fou, Yang:2020atz, Yang:2021izl} phenomenology. 

It is worth highlighting that the set of model parameters are fitted to reproduce a certain number of hadron observables within a given range of agreement with experiment. Therefore, it is difficult to assign an error to those parameters and, as a consequence, to the magnitudes calculated when using them. As the range of agreement between theory and experiment is around $10\%-20\%$, this value can be taken as an estimation of the model uncertainty.

The many-body Schr\"odinger equation including all the terms above  is solved by a diffusion Monte Carlo (DMC) technique which, in contrast with variational methods, considers in full the correlations between the particles of the system, and it is able to produce the exact energy of the system if we start with a reasonable initial approximation to the wave function.    


\noindent\emph{Theoretical formalism}.\,---\, Quantum Monte Carlo (QMC) methods have been successfully applied to many research areas such as quantum chemistry and material science~\cite{Hammond:1994bk, Foulkes:2001zz, Nightingale:2014bk}. The use of QMC methods to hadron physics has been scarce, basically because these methods are ideally suited to answer questions related with many-body physics and most known hadrons consist on 2- and 3-body bound states. This paradigm is changing in the last twenty years with many experimental signals indicating the possibility of having a zoo of tetra-, penta- and even hexa-quark systems~\cite{Zyla:2020zbs}.

Carlson \emph{et al.}~\cite{Carlson:1982xi, Carlson:1983rw} applied for the first time a Variational Quantum Monte Carlo (VQMC) algorithm, originally designed for nuclear physics problems, to the spectra of mesons and baryons. Their results compared reasonably well with those of the well-known Isgur-Karl's quark model~\cite{Isgur:1978xj, Isgur:1978wd, Isgur:1979be, Capstick:1986bm}. Since that exploratory work, there was almost no related activity until 2020 when a DMC algorithm was used to calculate the full spectrum of fully-heavy tetraquark system~\cite{Gordillo:2020sgc}. We follow the results of that work, and solved the Schr\"odinger equation that includes all the potential terms defined above using a Diffusion Monte Carlo technique for describing the isoscalar and isovector $J^{PC}=1^{++}$ $c\bar c n \bar n$ system.


\noindent\emph{Results}.\,---\, We obtain for the $J^{PC}=1^{++}$ $c \bar c n \bar n$ system, in the isoscalar and isovector sectors, the binding energies $-468\,\text{MeV}$ and $-460\,\text{MeV}$, which correspond to the absolute masses $3834\,\text{MeV}$ and $3842\,\text{MeV}$. Note that these masses are below the $D D^{\ast}$, $\rho J/\psi$ and $\omega J/\psi$ theoretical thresholds, located all at around $3870\,\text{MeV}$. As in Ref.~\cite{SilvestreBrac:1996bg}, the used quark masses are $m_u=315\,\text{MeV}$ and $m_c=1836\,\text{MeV}$; they can be fine-tuned in order to get agreement with the experimental mass of the $X(3872)$. In any case, the model uncertainty allows well to assert that theoretical and experimental masses are in fair agreement.

We are now interested on elucidating the structure of the two bound-states obtained above and thus we exploit the concept of radial distribution function because it provides valuable information of the existence of interquark correlations; in particular, $2$-body correlations. If the $n$-particle wave function is defined as $\psi(\vec{r}_1,\ldots,\vec{r}_n)$, where spin, flavor and color degrees of freedom have been ignored for simplicity without loss of generality, the probability of finding particle $1$ in position $\vec{r}_1$, particle $2$ in position $\vec{r}_2$, $\ldots$, particle $n$ in position $\vec{r}_n$ is:
\begin{equation}
P(\vec{r}_1,\ldots,\vec{r}_n) = \psi^{\ast}(\vec{r}_1,\ldots,\vec{r}_n) \psi(\vec{r}_1,\ldots,\vec{r}_n) \,,
\end{equation}
and it is normalize to one, \emph{i.e.}
\begin{equation}
1 = \int d\vec{r}_1\, \cdots d\vec{r}_n\, P(\vec{r}_1,\ldots,\vec{r}_n) \,.
\end{equation}
Therefore, one can define
\begin{equation}
\rho^{(2)}(\vec{r}_1,\vec{r}_2) = \int d\vec{r}_{3}\, \cdots d\vec{r}_{n}\,P(\vec{r}_1,\ldots,\vec{r}_n) \,,
\end{equation}
which expresses the probability of finding 2 particles in positions $\vec{r}_1$ and $\vec{r}_2$; and the radial distribution function as
\begin{equation}
\rho(r) = 4\pi r^2 \int d\vec{R}\, \rho^{(2)}(\vec{R}+\vec{r},\vec{R}) \,,
\end{equation}
where $r$ indicates now the distance between the two particles considered.

\begin{figure*}
\includegraphics[width=0.475\textwidth]{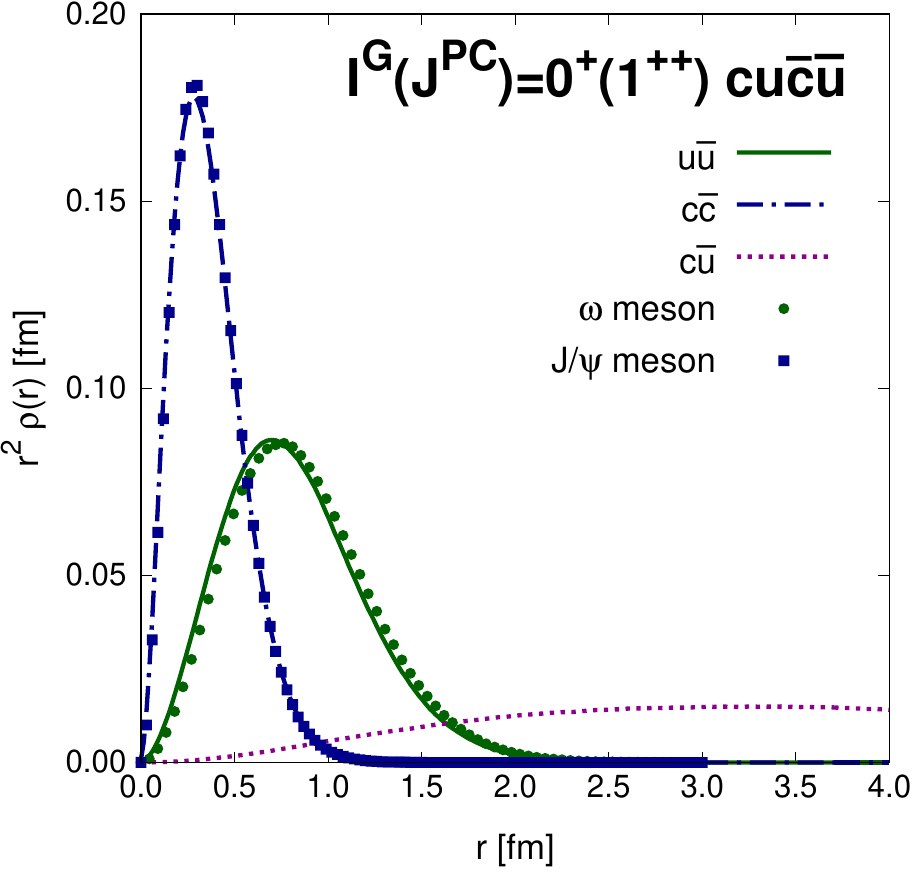}
\includegraphics[width=0.475\textwidth]{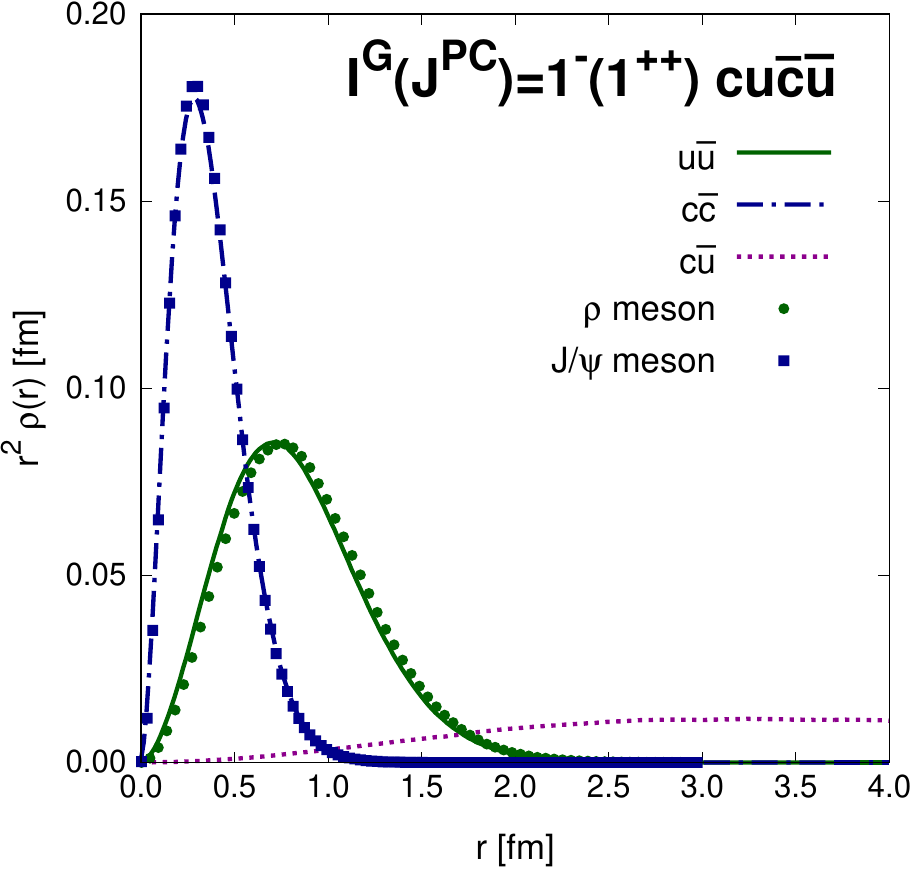}
\caption{\label{fig:Correlations} Radial distribution functions (see text for details) for the studied $X(3872)$ candidate as either $I^G(J^{PC})=0^+(1^{++})$ (left panel) or $I^G(J^{PC})=1^-(1^{++})$ (right panel) $c \bar c n \bar n$ tetraquark bound-state. These functions represent the probability of finding the 2 quarks (antiquarks) at an interquark distance $r$. In both panels, solid (green), dot-dashed (blue) and dotted (purple) represent, respectively, $n\bar n$, $c\bar c$ and $c\bar n$ correlations inside the $c \bar c n \bar n$ tetraquark. The dot (green) and square (blue) points stand for the same object but calculated for the corresponding mesons, \emph{i.e.} $\omega-J/\psi$ for the left panel and $\rho-J/\psi$ for the right panel.}
\end{figure*}

Figure~\ref{fig:Correlations} shows the radial distribution functions for the studied $X(3872)$ candidate as either $I^G(J^{PC})=0^+(1^{++})$ (left panel) or $I^G(J^{PC})=1^-(1^{++})$ (right panel) $c \bar c n \bar n$ tetraquark bound-state. These functions represent the probability of finding two quarks (antiquarks) at an interquark distance $r$. In both panels, solid (green), dot-dashed (blue) and dotted (purple) represent, respectively, $n\bar n$, $c\bar c$ and $c\bar n$ correlations inside the $cn\bar c\bar n$ tetraquark. The dot (green) and square (blue) points stand for the same object but calculated for the corresponding isolated mesons, \emph{i.e.} $\omega-J/\psi$ for the left panel and $\rho-J/\psi$ for the right panel.

Looking at the results on Fig.~\ref{fig:Correlations}, one can conclude: (i) the $J^{PC}=1^{++}$ $c \bar c n \bar n$ tetraquark tends to cluster in a meson-meson configuration and not in a diquark-antidiquark one; (ii) both quark-antiquark correlations have an extension $\lesssim1\,\text{fm}$, separated by a distance of around $3\,\text{fm}$; (iii) the preferred arrangement is the one for which the quarks appear as a $(n\bar n)-(c\bar c)$ pair of clusters, contrary to the general accepted $(n\bar c)-(c\bar n)$ one due to the closeness of the $X(3872)$'s mass to the $D^0\bar D^{\ast0}$ threshold; (iv) the $n\bar n$ correlation resembles closely the omega (rho) meson in the isospin zero (one) sector, the $c\bar c$ correlation is clearly a $J/\psi$ meson, note that all of them are ${}^3S_1$ quark-antiquark bound states. 

Some comments on the robustness of these results are due here. The Goldstone-boson exchange interactions between light quarks play a marginal role. For example, when turning them off, the binding energy of the tetraquark system changes by at most $8\%$. This indicates that the most general and accepted quark-(anti)quark interaction, Coulomb$\,+\,$linear-confining$\,+\,$hyperfine spin-spin, is behind the dynamics of the tetraquark. It is important to remind that the used chiral interaction has been fixed along two decades of studying hadron, hadron-hadron and multiquark phenomenology. When we performed an analysis of the $D^0\bar D^{\ast0}$ structure within this formalism and using, as a binding force, the same chiral interaction, a bound state is not produced probably because the interaction is too weak.

It should be mentioned that a recent study on the lowest-lying states of all-heavy tetraquark systems~\cite{Gordillo:2020sgc} sees a similar feature in the quark-quark correlations of the $J^{PC}=1^{++}$ $cb\bar c\bar b$ ground state (see left-bottom panel of Fig.~7 in Ref.~\cite{Gordillo:2020sgc}). The radial distribution functions of the $J^{PC}=1^{++}$ $cb\bar c\bar b$ ground state reveal that this state prefers to be organized in clusters of $c\bar c$ and $b\bar b$, whose extensions are less than $0.5\,\text{fm}$, separated by a distance larger than $0.8-1.0\,\text{fm}$. However, the arrangement of quarks (antiquarks) in the $J^{PC}=1^{++}$ $cb\bar c\bar b$ system is not repeated by its tetraquark partners with different quantum numbers $J^{PC}=0^{++}$, $1^{+-}$, and $2^{++}$; neither seen in any other case of fully-heavy tetraquarks explored in Ref.~\cite{Gordillo:2020sgc}: $cc\bar c \bar c$, $bb\bar b\bar b$, $cc\bar b\bar b$ ($bb\bar c\bar c$), $cc\bar c\bar b$, and $bb\bar c\bar b$. 

The DMC formalism introduced in Ref.~\cite{Gordillo:2020sgc} allows us to compute not only the eigenenergy but also its associated wave function. In this case, we obtain:\footnote{The space, spin and flavor components of the wave function are omitted without loss of generality.}
\begin{align}
|X(3872) \rangle_{\text{color}} = c_1 \, |\bar{3}_{nc} 3_{\bar n\bar c}\rangle_{\text{color}} + c_2 \, |6_{nc} \bar{6}_{\bar n \bar c}\rangle_{\text{color}} \,,
\end{align}
with $c_1\approx 0.57(1)$ and $c_2\approx 0.82(1)$, in both isoscalar and isovector $J^{PC}=1^{++}$ states. Looking at Eqs.~(56) and~(57) of Ref.~\cite{Gordillo:2020sgc}, our color wave function is essentially
\begin{align}
|X(3872) \rangle_{\text{color}} &\approx |1_{c\bar c} \bar{1}_{n\bar n} \rangle_{\text{color}} \,,
\end{align}
indicating that the computed tetraquark states are $(n\bar n)-(c\bar c)$ meson-meson configurations.\footnote{Within the same formalism, we have calculated the properties of the $|8_{c\bar c} \bar{8}_{n\bar n} \rangle_{\text{color}}$ state, which would be a color excitation. Its mass is around $150\,\text{MeV}$ above the ground state, and its radial distribution functions indicate that it is a compact object.} This supports the information related with the radial distribution functions shown in Fig.~\ref{fig:Correlations}, therein we report an extra piece of information, \emph{i.e.} such $n\bar n$ and $c\bar c$ correlations closely resemble either $\omega$ or $\rho$ mesons, in the light sector, and $J/\psi$ for the hidden-charm one.

Based on all the data above, our interpretation of the $X(3872)$ signal is that either two $c\bar c n\bar n$ bound states with quantum numbers $I^G(J^{PC})=0^+(1^{++})$ and $1^-(1^{++})$, respectively, or just the same two but coupled together may explain all the observed features of the $X(3872)$. Their masses appear close to the $D^0\bar D^{\ast0}$ threshold, a meson-meson molecular state is preferred but, contrary to the usual assumption of having $D^0\bar D^{\ast 0}$ molecule, our formalism produces $\omega J/\psi$ and $\rho J/\psi$ clusters which could explain (i) the $X(3872)$'s discovery decay channel $\pi^+\pi^-J/\psi$, despite violating isospin conservation; (ii) the ratio $R_{\omega-\rho}\approx 1$ measured by different experimental collaborations worldwide; (iii) the observed radiative decay rates $\gamma J/\psi$ and $\gamma \psi(2S)$, incompatible with $D^0\bar D^{\ast0}$-molecular interpretation, driven by the vector meson dominance mechanism, and (iv) production rates of the $X(3872)$ which are consistent with having a $c\bar c$ cluster.


\noindent\emph{Conclusions}.\,---\, We use a diffusion Monte Carlo method to solve the many-body Schr\"odinger equation that describes the $X(3872)$ as a $c \bar c n \bar n$ tetraquark system with quantum numbers $J^{PC}=1^{++}$. Among other advantages, this approach avoids the usual quark-clustering assumed in other theoretical techniques applied to the same problem and, moreover, provides information about the hadron's structural properties.

The interaction between particles was modeled by the most general and accepted potential, \emph{i.e.} a pairwise interaction including Coulomb, linear-confining and hyperfine spin-spin terms. There are also Goldstone-boson exchange interactions between light quarks that have been fixed in the last 10-20 years reproducing hadron, hadron-hadron and multiquark phenomenology. The chiral contribution to the mass of the $X(3872)$ represent at most $8\%$, leaving the rest for the general color interaction; note, too, that the chiral potentials are weak and they are not able to produce meson-meson molecular states.

We obtain two $c\bar c n \bar n$ bound states, with quantum numbers $I^G(J^{PC})=0^+(1^{++})$ and $1^-(1^{++})$, whose masses are below the relevant meson-meson thresholds. These states could contribute separately to the observed $X(3872)$ signal, or it may be the result of a coupling between them. According to our results, these states are arranged as light-meson--heavy-meson molecules of type $\omega J/\psi$ and $\rho J/\psi$, rather than the most extended $D^0\bar D^{\ast0}$ interpretation. This fact would be the key to make compatible the molecular features of the $X(3872)$ with its decay and production observables that seem to indicate the presence of a $c\bar c$ cluster. Finally, its multiplet partners do not show the same behavior, making the $J^{P}=1^{+}$ quantum numbers somewhat special, ideally suited to host molecules.


\noindent\emph{Acknowledgements}.\,---\, This work has been partially funded by the Ministerio Espa\~nol de Ciencia e Innovaci\'on under grant No. PID2019-107844GB-C22 and FIS2017-84114-C2-2-P; the Junta de Andaluc\'ia under contract No. Operativo FEDER Andaluc\'ia 2014-2020 UHU-1264517; but also PAIDI FQM-205 and -370. The authors acknowledges, too, the use of the computer facilities of C3UPO at the Universidad Pablo de Olavide, de Sevilla.


\bibliography{DMC-X3872}

\begin{thebibliography}{76}%
\makeatletter
\providecommand \@ifxundefined [1]{%
 \@ifx{#1\undefined}
}%
\providecommand \@ifnum [1]{%
 \ifnum #1\expandafter \@firstoftwo
 \else \expandafter \@secondoftwo
 \fi
}%
\providecommand \@ifx [1]{%
 \ifx #1\expandafter \@firstoftwo
 \else \expandafter \@secondoftwo
 \fi
}%
\providecommand \natexlab [1]{#1}%
\providecommand \enquote  [1]{``#1''}%
\providecommand \bibnamefont  [1]{#1}%
\providecommand \bibfnamefont [1]{#1}%
\providecommand \citenamefont [1]{#1}%
\providecommand \href@noop [0]{\@secondoftwo}%
\providecommand \href [0]{\begingroup \@sanitize@url \@href}%
\providecommand \@href[1]{\@@startlink{#1}\@@href}%
\providecommand \@@href[1]{\endgroup#1\@@endlink}%
\providecommand \@sanitize@url [0]{\catcode `\\12\catcode `\$12\catcode
  `\&12\catcode `\#12\catcode `\^12\catcode `\_12\catcode `\%12\relax}%
\providecommand \@@startlink[1]{}%
\providecommand \@@endlink[0]{}%
\providecommand \url  [0]{\begingroup\@sanitize@url \@url }%
\providecommand \@url [1]{\endgroup\@href {#1}{\urlprefix }}%
\providecommand \urlprefix  [0]{URL }%
\providecommand \Eprint [0]{\href }%
\providecommand \doibase [0]{http://dx.doi.org/}%
\providecommand \selectlanguage [0]{\@gobble}%
\providecommand \bibinfo  [0]{\@secondoftwo}%
\providecommand \bibfield  [0]{\@secondoftwo}%
\providecommand \translation [1]{[#1]}%
\providecommand \BibitemOpen [0]{}%
\providecommand \bibitemStop [0]{}%
\providecommand \bibitemNoStop [0]{.\EOS\space}%
\providecommand \EOS [0]{\spacefactor3000\relax}%
\providecommand \BibitemShut  [1]{\csname bibitem#1\endcsname}%
\let\auto@bib@innerbib\@empty
\bibitem [{\citenamefont {Gell-Mann}(1964)}]{GellMann:1964nj}%
  \BibitemOpen
  \bibfield  {author} {\bibinfo {author} {\bibfnamefont {M.}~\bibnamefont
  {Gell-Mann}},\ }\href {\doibase 10.1016/S0031-9163(64)92001-3} {\bibfield
  {journal} {\bibinfo  {journal} {Phys. Lett.}\ }\textbf {\bibinfo {volume}
  {8}},\ \bibinfo {pages} {214} (\bibinfo {year} {1964})}\BibitemShut {NoStop}%
\bibitem [{\citenamefont {Zweig}(1964)}]{Zweig:1964CERN}%
  \BibitemOpen
  \bibfield  {author} {\bibinfo {author} {\bibfnamefont {G.}~\bibnamefont
  {Zweig}},\ }\href@noop {} {\bibfield  {journal} {\bibinfo  {journal} {CERN
  Report No.8182/TH.401, CERN Report No.8419/TH.412}\ } (\bibinfo {year}
  {1964})}\BibitemShut {NoStop}%
\bibitem [{\citenamefont {Choi}\ \emph {et~al.}(2003)\citenamefont {Choi} \emph
  {et~al.}}]{Choi:2003ue}%
  \BibitemOpen
  \bibfield  {author} {\bibinfo {author} {\bibfnamefont {S.~K.}\ \bibnamefont
  {Choi}} \emph {et~al.} (\bibinfo {collaboration} {Belle}),\ }\href {\doibase
  10.1103/PhysRevLett.91.262001} {\bibfield  {journal} {\bibinfo  {journal}
  {Phys. Rev. Lett.}\ }\textbf {\bibinfo {volume} {91}},\ \bibinfo {pages}
  {262001} (\bibinfo {year} {2003})},\ \Eprint
  {http://arxiv.org/abs/hep-ex/0309032} {arXiv:hep-ex/0309032} \BibitemShut
  {NoStop}%
\bibitem [{\citenamefont {Lebed}\ \emph {et~al.}(2017)\citenamefont {Lebed},
  \citenamefont {Mitchell},\ and\ \citenamefont {Swanson}}]{Lebed:2016hpi}%
  \BibitemOpen
  \bibfield  {author} {\bibinfo {author} {\bibfnamefont {R.~F.}\ \bibnamefont
  {Lebed}}, \bibinfo {author} {\bibfnamefont {R.~E.}\ \bibnamefont {Mitchell}},
  \ and\ \bibinfo {author} {\bibfnamefont {E.~S.}\ \bibnamefont {Swanson}},\
  }\href {\doibase 10.1016/j.ppnp.2016.11.003} {\bibfield  {journal} {\bibinfo
  {journal} {Prog. Part. Nucl. Phys.}\ }\textbf {\bibinfo {volume} {93}},\
  \bibinfo {pages} {143} (\bibinfo {year} {2017})},\ \Eprint
  {http://arxiv.org/abs/1610.04528} {arXiv:1610.04528 [hep-ph]} \BibitemShut
  {NoStop}%
\bibitem [{\citenamefont {Ali}\ \emph {et~al.}(2017)\citenamefont {Ali},
  \citenamefont {Lange},\ and\ \citenamefont {Stone}}]{Ali:2017jda}%
  \BibitemOpen
  \bibfield  {author} {\bibinfo {author} {\bibfnamefont {A.}~\bibnamefont
  {Ali}}, \bibinfo {author} {\bibfnamefont {J.~S.}\ \bibnamefont {Lange}}, \
  and\ \bibinfo {author} {\bibfnamefont {S.}~\bibnamefont {Stone}},\ }\href
  {\doibase 10.1016/j.ppnp.2017.08.003} {\bibfield  {journal} {\bibinfo
  {journal} {Prog. Part. Nucl. Phys.}\ }\textbf {\bibinfo {volume} {97}},\
  \bibinfo {pages} {123} (\bibinfo {year} {2017})},\ \Eprint
  {http://arxiv.org/abs/1706.00610} {arXiv:1706.00610 [hep-ph]} \BibitemShut
  {NoStop}%
\bibitem [{\citenamefont {Guo}\ \emph {et~al.}(2018)\citenamefont {Guo},
  \citenamefont {Hanhart}, \citenamefont {Mei\ss{}ner}, \citenamefont {Wang},
  \citenamefont {Zhao},\ and\ \citenamefont {Zou}}]{Guo:2017jvc}%
  \BibitemOpen
  \bibfield  {author} {\bibinfo {author} {\bibfnamefont {F.-K.}\ \bibnamefont
  {Guo}}, \bibinfo {author} {\bibfnamefont {C.}~\bibnamefont {Hanhart}},
  \bibinfo {author} {\bibfnamefont {U.-G.}\ \bibnamefont {Mei\ss{}ner}},
  \bibinfo {author} {\bibfnamefont {Q.}~\bibnamefont {Wang}}, \bibinfo {author}
  {\bibfnamefont {Q.}~\bibnamefont {Zhao}}, \ and\ \bibinfo {author}
  {\bibfnamefont {B.-S.}\ \bibnamefont {Zou}},\ }\href {\doibase
  10.1103/RevModPhys.90.015004} {\bibfield  {journal} {\bibinfo  {journal}
  {Rev. Mod. Phys.}\ }\textbf {\bibinfo {volume} {90}},\ \bibinfo {pages}
  {015004} (\bibinfo {year} {2018})},\ \Eprint
  {http://arxiv.org/abs/1705.00141} {arXiv:1705.00141 [hep-ph]} \BibitemShut
  {NoStop}%
\bibitem [{\citenamefont {Olsen}\ \emph {et~al.}(2018)\citenamefont {Olsen},
  \citenamefont {Skwarnicki},\ and\ \citenamefont {Zieminska}}]{Olsen:2017bmm}%
  \BibitemOpen
  \bibfield  {author} {\bibinfo {author} {\bibfnamefont {S.~L.}\ \bibnamefont
  {Olsen}}, \bibinfo {author} {\bibfnamefont {T.}~\bibnamefont {Skwarnicki}}, \
  and\ \bibinfo {author} {\bibfnamefont {D.}~\bibnamefont {Zieminska}},\ }\href
  {\doibase 10.1103/RevModPhys.90.015003} {\bibfield  {journal} {\bibinfo
  {journal} {Rev. Mod. Phys.}\ }\textbf {\bibinfo {volume} {90}},\ \bibinfo
  {pages} {015003} (\bibinfo {year} {2018})},\ \Eprint
  {http://arxiv.org/abs/1708.04012} {arXiv:1708.04012 [hep-ph]} \BibitemShut
  {NoStop}%
\bibitem [{\citenamefont {Liu}\ \emph {et~al.}(2019)\citenamefont {Liu},
  \citenamefont {Chen}, \citenamefont {Chen}, \citenamefont {Liu},\ and\
  \citenamefont {Zhu}}]{Liu:2019zoy}%
  \BibitemOpen
  \bibfield  {author} {\bibinfo {author} {\bibfnamefont {Y.-R.}\ \bibnamefont
  {Liu}}, \bibinfo {author} {\bibfnamefont {H.-X.}\ \bibnamefont {Chen}},
  \bibinfo {author} {\bibfnamefont {W.}~\bibnamefont {Chen}}, \bibinfo {author}
  {\bibfnamefont {X.}~\bibnamefont {Liu}}, \ and\ \bibinfo {author}
  {\bibfnamefont {S.-L.}\ \bibnamefont {Zhu}},\ }\href {\doibase
  10.1016/j.ppnp.2019.04.003} {\bibfield  {journal} {\bibinfo  {journal} {Prog.
  Part. Nucl. Phys.}\ }\textbf {\bibinfo {volume} {107}},\ \bibinfo {pages}
  {237} (\bibinfo {year} {2019})},\ \Eprint {http://arxiv.org/abs/1903.11976}
  {arXiv:1903.11976 [hep-ph]} \BibitemShut {NoStop}%
\bibitem [{\citenamefont {Brambilla}\ \emph {et~al.}(2020)\citenamefont
  {Brambilla}, \citenamefont {Eidelman}, \citenamefont {Hanhart}, \citenamefont
  {Nefediev}, \citenamefont {Shen}, \citenamefont {Thomas}, \citenamefont
  {Vairo},\ and\ \citenamefont {Yuan}}]{Brambilla:2019esw}%
  \BibitemOpen
  \bibfield  {author} {\bibinfo {author} {\bibfnamefont {N.}~\bibnamefont
  {Brambilla}}, \bibinfo {author} {\bibfnamefont {S.}~\bibnamefont {Eidelman}},
  \bibinfo {author} {\bibfnamefont {C.}~\bibnamefont {Hanhart}}, \bibinfo
  {author} {\bibfnamefont {A.}~\bibnamefont {Nefediev}}, \bibinfo {author}
  {\bibfnamefont {C.-P.}\ \bibnamefont {Shen}}, \bibinfo {author}
  {\bibfnamefont {C.~E.}\ \bibnamefont {Thomas}}, \bibinfo {author}
  {\bibfnamefont {A.}~\bibnamefont {Vairo}}, \ and\ \bibinfo {author}
  {\bibfnamefont {C.-Z.}\ \bibnamefont {Yuan}},\ }\href {\doibase
  10.1016/j.physrep.2020.05.001} {\bibfield  {journal} {\bibinfo  {journal}
  {Phys. Rept.}\ }\textbf {\bibinfo {volume} {873}},\ \bibinfo {pages} {1}
  (\bibinfo {year} {2020})},\ \Eprint {http://arxiv.org/abs/1907.07583}
  {arXiv:1907.07583 [hep-ex]} \BibitemShut {NoStop}%
\bibitem [{\citenamefont {Zyla}\ \emph {et~al.}(2020)\citenamefont {Zyla} \emph
  {et~al.}}]{Zyla:2020zbs}%
  \BibitemOpen
  \bibfield  {author} {\bibinfo {author} {\bibfnamefont {P.~A.}\ \bibnamefont
  {Zyla}} \emph {et~al.} (\bibinfo {collaboration} {Particle Data Group}),\
  }\href {\doibase 10.1093/ptep/ptaa104} {\bibfield  {journal} {\bibinfo
  {journal} {PTEP}\ }\textbf {\bibinfo {volume} {2020}},\ \bibinfo {pages}
  {083C01} (\bibinfo {year} {2020})}\BibitemShut {NoStop}%
\bibitem [{\citenamefont {Abe}\ \emph {et~al.}(2005{\natexlab{a}})\citenamefont
  {Abe} \emph {et~al.}}]{Abe:2005ix}%
  \BibitemOpen
  \bibfield  {author} {\bibinfo {author} {\bibfnamefont {K.}~\bibnamefont
  {Abe}} \emph {et~al.} (\bibinfo {collaboration} {Belle}),\ }in\ \href@noop {}
  {\emph {\bibinfo {booktitle} {{22nd International Symposium on Lepton-Photon
  Interactions at High Energy (LP 2005)}}}}\ (\bibinfo {year} {2005})\ \Eprint
  {http://arxiv.org/abs/hep-ex/0505037} {arXiv:hep-ex/0505037} \BibitemShut
  {NoStop}%
\bibitem [{\citenamefont {Abe}\ \emph {et~al.}(2005{\natexlab{b}})\citenamefont
  {Abe} \emph {et~al.}}]{Abe:2005iya}%
  \BibitemOpen
  \bibfield  {author} {\bibinfo {author} {\bibfnamefont {K.}~\bibnamefont
  {Abe}} \emph {et~al.} (\bibinfo {collaboration} {Belle}),\ }in\ \href@noop {}
  {\emph {\bibinfo {booktitle} {{22nd International Symposium on Lepton-Photon
  Interactions at High Energy (LP 2005)}}}}\ (\bibinfo {year} {2005})\ \Eprint
  {http://arxiv.org/abs/hep-ex/0505038} {arXiv:hep-ex/0505038} \BibitemShut
  {NoStop}%
\bibitem [{\citenamefont {Abulencia}\ \emph {et~al.}(2006)\citenamefont
  {Abulencia} \emph {et~al.}}]{Abulencia:2005zc}%
  \BibitemOpen
  \bibfield  {author} {\bibinfo {author} {\bibfnamefont {A.}~\bibnamefont
  {Abulencia}} \emph {et~al.} (\bibinfo {collaboration} {CDF}),\ }\href
  {\doibase 10.1103/PhysRevLett.96.102002} {\bibfield  {journal} {\bibinfo
  {journal} {Phys. Rev. Lett.}\ }\textbf {\bibinfo {volume} {96}},\ \bibinfo
  {pages} {102002} (\bibinfo {year} {2006})},\ \Eprint
  {http://arxiv.org/abs/hep-ex/0512074} {arXiv:hep-ex/0512074} \BibitemShut
  {NoStop}%
\bibitem [{\citenamefont {Abulencia}\ \emph {et~al.}(2007)\citenamefont
  {Abulencia} \emph {et~al.}}]{Abulencia:2006ma}%
  \BibitemOpen
  \bibfield  {author} {\bibinfo {author} {\bibfnamefont {A.}~\bibnamefont
  {Abulencia}} \emph {et~al.} (\bibinfo {collaboration} {CDF}),\ }\href
  {\doibase 10.1103/PhysRevLett.98.132002} {\bibfield  {journal} {\bibinfo
  {journal} {Phys. Rev. Lett.}\ }\textbf {\bibinfo {volume} {98}},\ \bibinfo
  {pages} {132002} (\bibinfo {year} {2007})},\ \Eprint
  {http://arxiv.org/abs/hep-ex/0612053} {arXiv:hep-ex/0612053} \BibitemShut
  {NoStop}%
\bibitem [{\citenamefont {Ebert}\ \emph {et~al.}(2003)\citenamefont {Ebert},
  \citenamefont {Faustov},\ and\ \citenamefont {Galkin}}]{Ebert:2002pp}%
  \BibitemOpen
  \bibfield  {author} {\bibinfo {author} {\bibfnamefont {D.}~\bibnamefont
  {Ebert}}, \bibinfo {author} {\bibfnamefont {R.~N.}\ \bibnamefont {Faustov}},
  \ and\ \bibinfo {author} {\bibfnamefont {V.~O.}\ \bibnamefont {Galkin}},\
  }\href {\doibase 10.1103/PhysRevD.67.014027} {\bibfield  {journal} {\bibinfo
  {journal} {Phys. Rev. D}\ }\textbf {\bibinfo {volume} {67}},\ \bibinfo
  {pages} {014027} (\bibinfo {year} {2003})},\ \Eprint
  {http://arxiv.org/abs/hep-ph/0210381} {arXiv:hep-ph/0210381} \BibitemShut
  {NoStop}%
\bibitem [{\citenamefont {Barnes}\ \emph {et~al.}(2005)\citenamefont {Barnes},
  \citenamefont {Godfrey},\ and\ \citenamefont {Swanson}}]{Barnes:2005pb}%
  \BibitemOpen
  \bibfield  {author} {\bibinfo {author} {\bibfnamefont {T.}~\bibnamefont
  {Barnes}}, \bibinfo {author} {\bibfnamefont {S.}~\bibnamefont {Godfrey}}, \
  and\ \bibinfo {author} {\bibfnamefont {E.~S.}\ \bibnamefont {Swanson}},\
  }\href {\doibase 10.1103/PhysRevD.72.054026} {\bibfield  {journal} {\bibinfo
  {journal} {Phys. Rev. D}\ }\textbf {\bibinfo {volume} {72}},\ \bibinfo
  {pages} {054026} (\bibinfo {year} {2005})},\ \Eprint
  {http://arxiv.org/abs/hep-ph/0505002} {arXiv:hep-ph/0505002} \BibitemShut
  {NoStop}%
\bibitem [{\citenamefont {Segovia}\ \emph {et~al.}(2013)\citenamefont
  {Segovia}, \citenamefont {Entem}, \citenamefont {Fernandez},\ and\
  \citenamefont {Hernandez}}]{Segovia:2013wma}%
  \BibitemOpen
  \bibfield  {author} {\bibinfo {author} {\bibfnamefont {J.}~\bibnamefont
  {Segovia}}, \bibinfo {author} {\bibfnamefont {D.~R.}\ \bibnamefont {Entem}},
  \bibinfo {author} {\bibfnamefont {F.}~\bibnamefont {Fernandez}}, \ and\
  \bibinfo {author} {\bibfnamefont {E.}~\bibnamefont {Hernandez}},\ }\href
  {\doibase 10.1142/S0218301313300269} {\bibfield  {journal} {\bibinfo
  {journal} {Int. J. Mod. Phys. E}\ }\textbf {\bibinfo {volume} {22}},\
  \bibinfo {pages} {1330026} (\bibinfo {year} {2013})},\ \Eprint
  {http://arxiv.org/abs/1309.6926} {arXiv:1309.6926 [hep-ph]} \BibitemShut
  {NoStop}%
\bibitem [{\citenamefont {Tornqvist}(1994)}]{Tornqvist:1993ng}%
  \BibitemOpen
  \bibfield  {author} {\bibinfo {author} {\bibfnamefont {N.~A.}\ \bibnamefont
  {Tornqvist}},\ }\href {\doibase 10.1007/BF01413192} {\bibfield  {journal}
  {\bibinfo  {journal} {Z. Phys. C}\ }\textbf {\bibinfo {volume} {61}},\
  \bibinfo {pages} {525} (\bibinfo {year} {1994})},\ \Eprint
  {http://arxiv.org/abs/hep-ph/9310247} {arXiv:hep-ph/9310247} \BibitemShut
  {NoStop}%
\bibitem [{\citenamefont {Close}\ and\ \citenamefont
  {Page}(2004)}]{Close:2003sg}%
  \BibitemOpen
  \bibfield  {author} {\bibinfo {author} {\bibfnamefont {F.~E.}\ \bibnamefont
  {Close}}\ and\ \bibinfo {author} {\bibfnamefont {P.~R.}\ \bibnamefont
  {Page}},\ }\href {\doibase 10.1016/j.physletb.2003.10.032} {\bibfield
  {journal} {\bibinfo  {journal} {Phys. Lett. B}\ }\textbf {\bibinfo {volume}
  {578}},\ \bibinfo {pages} {119} (\bibinfo {year} {2004})},\ \Eprint
  {http://arxiv.org/abs/hep-ph/0309253} {arXiv:hep-ph/0309253} \BibitemShut
  {NoStop}%
\bibitem [{\citenamefont {Swanson}(2004)}]{Swanson:2003tb}%
  \BibitemOpen
  \bibfield  {author} {\bibinfo {author} {\bibfnamefont {E.~S.}\ \bibnamefont
  {Swanson}},\ }\href {\doibase 10.1016/j.physletb.2004.03.033} {\bibfield
  {journal} {\bibinfo  {journal} {Phys. Lett. B}\ }\textbf {\bibinfo {volume}
  {588}},\ \bibinfo {pages} {189} (\bibinfo {year} {2004})},\ \Eprint
  {http://arxiv.org/abs/hep-ph/0311229} {arXiv:hep-ph/0311229} \BibitemShut
  {NoStop}%
\bibitem [{\citenamefont {Voloshin}(2004)}]{Voloshin:2003nt}%
  \BibitemOpen
  \bibfield  {author} {\bibinfo {author} {\bibfnamefont {M.~B.}\ \bibnamefont
  {Voloshin}},\ }\href {\doibase 10.1016/j.physletb.2003.11.014} {\bibfield
  {journal} {\bibinfo  {journal} {Phys. Lett. B}\ }\textbf {\bibinfo {volume}
  {579}},\ \bibinfo {pages} {316} (\bibinfo {year} {2004})},\ \Eprint
  {http://arxiv.org/abs/hep-ph/0309307} {arXiv:hep-ph/0309307} \BibitemShut
  {NoStop}%
\bibitem [{\citenamefont {Wong}(2004)}]{Wong:2003xk}%
  \BibitemOpen
  \bibfield  {author} {\bibinfo {author} {\bibfnamefont {C.-Y.}\ \bibnamefont
  {Wong}},\ }\href {\doibase 10.1103/PhysRevC.69.055202} {\bibfield  {journal}
  {\bibinfo  {journal} {Phys. Rev. C}\ }\textbf {\bibinfo {volume} {69}},\
  \bibinfo {pages} {055202} (\bibinfo {year} {2004})},\ \Eprint
  {http://arxiv.org/abs/hep-ph/0311088} {arXiv:hep-ph/0311088} \BibitemShut
  {NoStop}%
\bibitem [{\citenamefont {Fleming}\ \emph {et~al.}(2007)\citenamefont
  {Fleming}, \citenamefont {Kusunoki}, \citenamefont {Mehen},\ and\
  \citenamefont {van Kolck}}]{Fleming:2007rp}%
  \BibitemOpen
  \bibfield  {author} {\bibinfo {author} {\bibfnamefont {S.}~\bibnamefont
  {Fleming}}, \bibinfo {author} {\bibfnamefont {M.}~\bibnamefont {Kusunoki}},
  \bibinfo {author} {\bibfnamefont {T.}~\bibnamefont {Mehen}}, \ and\ \bibinfo
  {author} {\bibfnamefont {U.}~\bibnamefont {van Kolck}},\ }\href {\doibase
  10.1103/PhysRevD.76.034006} {\bibfield  {journal} {\bibinfo  {journal} {Phys.
  Rev. D}\ }\textbf {\bibinfo {volume} {76}},\ \bibinfo {pages} {034006}
  (\bibinfo {year} {2007})},\ \Eprint {http://arxiv.org/abs/hep-ph/0703168}
  {arXiv:hep-ph/0703168} \BibitemShut {NoStop}%
\bibitem [{\citenamefont {del Amo~Sanchez}\ \emph {et~al.}(2010)\citenamefont
  {del Amo~Sanchez} \emph {et~al.}}]{delAmoSanchez:2010jr}%
  \BibitemOpen
  \bibfield  {author} {\bibinfo {author} {\bibfnamefont {P.}~\bibnamefont {del
  Amo~Sanchez}} \emph {et~al.} (\bibinfo {collaboration} {BaBar}),\ }\href
  {\doibase 10.1103/PhysRevD.82.011101} {\bibfield  {journal} {\bibinfo
  {journal} {Phys. Rev. D}\ }\textbf {\bibinfo {volume} {82}},\ \bibinfo
  {pages} {011101} (\bibinfo {year} {2010})},\ \Eprint
  {http://arxiv.org/abs/1005.5190} {arXiv:1005.5190 [hep-ex]} \BibitemShut
  {NoStop}%
\bibitem [{\citenamefont {Ablikim}\ \emph {et~al.}(2019)\citenamefont {Ablikim}
  \emph {et~al.}}]{Ablikim:2019zio}%
  \BibitemOpen
  \bibfield  {author} {\bibinfo {author} {\bibfnamefont {M.}~\bibnamefont
  {Ablikim}} \emph {et~al.} (\bibinfo {collaboration} {BESIII}),\ }\href
  {\doibase 10.1103/PhysRevLett.122.232002} {\bibfield  {journal} {\bibinfo
  {journal} {Phys. Rev. Lett.}\ }\textbf {\bibinfo {volume} {122}},\ \bibinfo
  {pages} {232002} (\bibinfo {year} {2019})},\ \Eprint
  {http://arxiv.org/abs/1903.04695} {arXiv:1903.04695 [hep-ex]} \BibitemShut
  {NoStop}%
\bibitem [{\citenamefont {Matheus}\ \emph {et~al.}(2009)\citenamefont
  {Matheus}, \citenamefont {Navarra}, \citenamefont {Nielsen},\ and\
  \citenamefont {Zanetti}}]{Matheus:2009vq}%
  \BibitemOpen
  \bibfield  {author} {\bibinfo {author} {\bibfnamefont {R.~D.}\ \bibnamefont
  {Matheus}}, \bibinfo {author} {\bibfnamefont {F.~S.}\ \bibnamefont
  {Navarra}}, \bibinfo {author} {\bibfnamefont {M.}~\bibnamefont {Nielsen}}, \
  and\ \bibinfo {author} {\bibfnamefont {C.~M.}\ \bibnamefont {Zanetti}},\
  }\href {\doibase 10.1103/PhysRevD.80.056002} {\bibfield  {journal} {\bibinfo
  {journal} {Phys. Rev. D}\ }\textbf {\bibinfo {volume} {80}},\ \bibinfo
  {pages} {056002} (\bibinfo {year} {2009})},\ \Eprint
  {http://arxiv.org/abs/0907.2683} {arXiv:0907.2683 [hep-ph]} \BibitemShut
  {NoStop}%
\bibitem [{\citenamefont {Eichten}\ \emph {et~al.}(2004)\citenamefont
  {Eichten}, \citenamefont {Lane},\ and\ \citenamefont
  {Quigg}}]{Eichten:2004uh}%
  \BibitemOpen
  \bibfield  {author} {\bibinfo {author} {\bibfnamefont {E.~J.}\ \bibnamefont
  {Eichten}}, \bibinfo {author} {\bibfnamefont {K.}~\bibnamefont {Lane}}, \
  and\ \bibinfo {author} {\bibfnamefont {C.}~\bibnamefont {Quigg}},\ }\href
  {\doibase 10.1103/PhysRevD.69.094019} {\bibfield  {journal} {\bibinfo
  {journal} {Phys. Rev. D}\ }\textbf {\bibinfo {volume} {69}},\ \bibinfo
  {pages} {094019} (\bibinfo {year} {2004})},\ \Eprint
  {http://arxiv.org/abs/hep-ph/0401210} {arXiv:hep-ph/0401210} \BibitemShut
  {NoStop}%
\bibitem [{\citenamefont {Suzuki}(2005)}]{Suzuki:2005ha}%
  \BibitemOpen
  \bibfield  {author} {\bibinfo {author} {\bibfnamefont {M.}~\bibnamefont
  {Suzuki}},\ }\href {\doibase 10.1103/PhysRevD.72.114013} {\bibfield
  {journal} {\bibinfo  {journal} {Phys. Rev. D}\ }\textbf {\bibinfo {volume}
  {72}},\ \bibinfo {pages} {114013} (\bibinfo {year} {2005})},\ \Eprint
  {http://arxiv.org/abs/hep-ph/0508258} {arXiv:hep-ph/0508258} \BibitemShut
  {NoStop}%
\bibitem [{\citenamefont {Dong}\ \emph {et~al.}(2008)\citenamefont {Dong},
  \citenamefont {Faessler}, \citenamefont {Gutsche},\ and\ \citenamefont
  {Lyubovitskij}}]{Dong:2008gb}%
  \BibitemOpen
  \bibfield  {author} {\bibinfo {author} {\bibfnamefont {Y.-b.}\ \bibnamefont
  {Dong}}, \bibinfo {author} {\bibfnamefont {A.}~\bibnamefont {Faessler}},
  \bibinfo {author} {\bibfnamefont {T.}~\bibnamefont {Gutsche}}, \ and\
  \bibinfo {author} {\bibfnamefont {V.~E.}\ \bibnamefont {Lyubovitskij}},\
  }\href {\doibase 10.1103/PhysRevD.77.094013} {\bibfield  {journal} {\bibinfo
  {journal} {Phys. Rev. D}\ }\textbf {\bibinfo {volume} {77}},\ \bibinfo
  {pages} {094013} (\bibinfo {year} {2008})},\ \Eprint
  {http://arxiv.org/abs/0802.3610} {arXiv:0802.3610 [hep-ph]} \BibitemShut
  {NoStop}%
\bibitem [{\citenamefont {Ortega}\ \emph {et~al.}(2010)\citenamefont {Ortega},
  \citenamefont {Segovia}, \citenamefont {Entem},\ and\ \citenamefont
  {Fernandez}}]{Ortega:2009hj}%
  \BibitemOpen
  \bibfield  {author} {\bibinfo {author} {\bibfnamefont {P.~G.}\ \bibnamefont
  {Ortega}}, \bibinfo {author} {\bibfnamefont {J.}~\bibnamefont {Segovia}},
  \bibinfo {author} {\bibfnamefont {D.~R.}\ \bibnamefont {Entem}}, \ and\
  \bibinfo {author} {\bibfnamefont {F.}~\bibnamefont {Fernandez}},\ }\href
  {\doibase 10.1103/PhysRevD.81.054023} {\bibfield  {journal} {\bibinfo
  {journal} {Phys. Rev. D}\ }\textbf {\bibinfo {volume} {81}},\ \bibinfo
  {pages} {054023} (\bibinfo {year} {2010})},\ \Eprint
  {http://arxiv.org/abs/0907.3997} {arXiv:0907.3997 [hep-ph]} \BibitemShut
  {NoStop}%
\bibitem [{\citenamefont {Maiani}\ \emph {et~al.}(2005)\citenamefont {Maiani},
  \citenamefont {Piccinini}, \citenamefont {Polosa},\ and\ \citenamefont
  {Riquer}}]{Maiani:2004vq}%
  \BibitemOpen
  \bibfield  {author} {\bibinfo {author} {\bibfnamefont {L.}~\bibnamefont
  {Maiani}}, \bibinfo {author} {\bibfnamefont {F.}~\bibnamefont {Piccinini}},
  \bibinfo {author} {\bibfnamefont {A.~D.}\ \bibnamefont {Polosa}}, \ and\
  \bibinfo {author} {\bibfnamefont {V.}~\bibnamefont {Riquer}},\ }\href
  {\doibase 10.1103/PhysRevD.71.014028} {\bibfield  {journal} {\bibinfo
  {journal} {Phys. Rev. D}\ }\textbf {\bibinfo {volume} {71}},\ \bibinfo
  {pages} {014028} (\bibinfo {year} {2005})},\ \Eprint
  {http://arxiv.org/abs/hep-ph/0412098} {arXiv:hep-ph/0412098} \BibitemShut
  {NoStop}%
\bibitem [{\citenamefont {Anselmino}\ \emph {et~al.}(1993)\citenamefont
  {Anselmino}, \citenamefont {Predazzi}, \citenamefont {Ekelin}, \citenamefont
  {Fredriksson},\ and\ \citenamefont {Lichtenberg}}]{Anselmino:1992vg}%
  \BibitemOpen
  \bibfield  {author} {\bibinfo {author} {\bibfnamefont {M.}~\bibnamefont
  {Anselmino}}, \bibinfo {author} {\bibfnamefont {E.}~\bibnamefont {Predazzi}},
  \bibinfo {author} {\bibfnamefont {S.}~\bibnamefont {Ekelin}}, \bibinfo
  {author} {\bibfnamefont {S.}~\bibnamefont {Fredriksson}}, \ and\ \bibinfo
  {author} {\bibfnamefont {D.~B.}\ \bibnamefont {Lichtenberg}},\ }\href
  {\doibase 10.1103/RevModPhys.65.1199} {\bibfield  {journal} {\bibinfo
  {journal} {Rev. Mod. Phys.}\ }\textbf {\bibinfo {volume} {65}},\ \bibinfo
  {pages} {1199} (\bibinfo {year} {1993})}\BibitemShut {NoStop}%
\bibitem [{\citenamefont {Close}(2005)}]{Close:2004ip}%
  \BibitemOpen
  \bibfield  {author} {\bibinfo {author} {\bibfnamefont {F.~E.}\ \bibnamefont
  {Close}},\ }\href {\doibase 10.1142/S0217751X05028661} {\bibfield  {journal}
  {\bibinfo  {journal} {Int. J. Mod. Phys. A}\ }\textbf {\bibinfo {volume}
  {20}},\ \bibinfo {pages} {5156} (\bibinfo {year} {2005})},\ \Eprint
  {http://arxiv.org/abs/hep-ph/0411396} {arXiv:hep-ph/0411396} \BibitemShut
  {NoStop}%
\bibitem [{\citenamefont {Selem}\ and\ \citenamefont
  {Wilczek}(2006)}]{Selem:2006nd}%
  \BibitemOpen
  \bibfield  {author} {\bibinfo {author} {\bibfnamefont {A.}~\bibnamefont
  {Selem}}\ and\ \bibinfo {author} {\bibfnamefont {F.}~\bibnamefont
  {Wilczek}},\ }in\ \href {\doibase 10.1142/9789812773524_0030} {\emph
  {\bibinfo {booktitle} {{Ringberg Workshop on New Trends in HERA Physics
  2005}}}}\ (\bibinfo {year} {2006})\ \Eprint
  {http://arxiv.org/abs/hep-ph/0602128} {arXiv:hep-ph/0602128} \BibitemShut
  {NoStop}%
\bibitem [{\citenamefont {Friedmann}(2013)}]{Friedmann:2009mx}%
  \BibitemOpen
  \bibfield  {author} {\bibinfo {author} {\bibfnamefont {T.}~\bibnamefont
  {Friedmann}},\ }\href {\doibase 10.1140/epjc/s10052-013-2298-9} {\bibfield
  {journal} {\bibinfo  {journal} {Eur. Phys. J. C}\ }\textbf {\bibinfo {volume}
  {73}},\ \bibinfo {pages} {2298} (\bibinfo {year} {2013})},\ \Eprint
  {http://arxiv.org/abs/0910.2229} {arXiv:0910.2229 [hep-ph]} \BibitemShut
  {NoStop}%
\bibitem [{\citenamefont {Barabanov}\ \emph {et~al.}(2021)\citenamefont
  {Barabanov} \emph {et~al.}}]{Barabanov:2020jvn}%
  \BibitemOpen
  \bibfield  {author} {\bibinfo {author} {\bibfnamefont {M.~Y.}\ \bibnamefont
  {Barabanov}} \emph {et~al.},\ }\href {\doibase 10.1016/j.ppnp.2020.103835}
  {\bibfield  {journal} {\bibinfo  {journal} {Prog. Part. Nucl. Phys.}\
  }\textbf {\bibinfo {volume} {116}},\ \bibinfo {pages} {103835} (\bibinfo
  {year} {2021})},\ \Eprint {http://arxiv.org/abs/2008.07630} {arXiv:2008.07630
  [hep-ph]} \BibitemShut {NoStop}%
\bibitem [{\citenamefont {Drenska}\ \emph {et~al.}(2008)\citenamefont
  {Drenska}, \citenamefont {Faccini},\ and\ \citenamefont
  {Polosa}}]{Drenska:2008gr}%
  \BibitemOpen
  \bibfield  {author} {\bibinfo {author} {\bibfnamefont {N.~V.}\ \bibnamefont
  {Drenska}}, \bibinfo {author} {\bibfnamefont {R.}~\bibnamefont {Faccini}}, \
  and\ \bibinfo {author} {\bibfnamefont {A.~D.}\ \bibnamefont {Polosa}},\
  }\href {\doibase 10.1016/j.physletb.2008.09.038} {\bibfield  {journal}
  {\bibinfo  {journal} {Phys. Lett. B}\ }\textbf {\bibinfo {volume} {669}},\
  \bibinfo {pages} {160} (\bibinfo {year} {2008})},\ \Eprint
  {http://arxiv.org/abs/0807.0593} {arXiv:0807.0593 [hep-ph]} \BibitemShut
  {NoStop}%
\bibitem [{\citenamefont {Drenska}\ \emph {et~al.}(2009)\citenamefont
  {Drenska}, \citenamefont {Faccini},\ and\ \citenamefont
  {Polosa}}]{Drenska:2009cd}%
  \BibitemOpen
  \bibfield  {author} {\bibinfo {author} {\bibfnamefont {N.~V.}\ \bibnamefont
  {Drenska}}, \bibinfo {author} {\bibfnamefont {R.}~\bibnamefont {Faccini}}, \
  and\ \bibinfo {author} {\bibfnamefont {A.~D.}\ \bibnamefont {Polosa}},\
  }\href {\doibase 10.1103/PhysRevD.79.077502} {\bibfield  {journal} {\bibinfo
  {journal} {Phys. Rev. D}\ }\textbf {\bibinfo {volume} {79}},\ \bibinfo
  {pages} {077502} (\bibinfo {year} {2009})},\ \Eprint
  {http://arxiv.org/abs/0902.2803} {arXiv:0902.2803 [hep-ph]} \BibitemShut
  {NoStop}%
\bibitem [{\citenamefont {Semay}\ and\ \citenamefont
  {Silvestre-Brac}(1994)}]{Semay:1994ht}%
  \BibitemOpen
  \bibfield  {author} {\bibinfo {author} {\bibfnamefont {C.}~\bibnamefont
  {Semay}}\ and\ \bibinfo {author} {\bibfnamefont {B.}~\bibnamefont
  {Silvestre-Brac}},\ }\href {\doibase 10.1007/BF01413104} {\bibfield
  {journal} {\bibinfo  {journal} {Z. Phys. C}\ }\textbf {\bibinfo {volume}
  {61}},\ \bibinfo {pages} {271} (\bibinfo {year} {1994})}\BibitemShut
  {NoStop}%
\bibitem [{\citenamefont {Silvestre-Brac}(1996)}]{SilvestreBrac:1996bg}%
  \BibitemOpen
  \bibfield  {author} {\bibinfo {author} {\bibfnamefont {B.}~\bibnamefont
  {Silvestre-Brac}},\ }\href {\doibase 10.1007/s006010050028} {\bibfield
  {journal} {\bibinfo  {journal} {Few Body Syst.}\ }\textbf {\bibinfo {volume}
  {20}},\ \bibinfo {pages} {1} (\bibinfo {year} {1996})}\BibitemShut {NoStop}%
\bibitem [{\citenamefont {Diakonov}(2003)}]{Diakonov:2002fq}%
  \BibitemOpen
  \bibfield  {author} {\bibinfo {author} {\bibfnamefont {D.}~\bibnamefont
  {Diakonov}},\ }\href {\doibase 10.1016/S0146-6410(03)90014-7} {\bibfield
  {journal} {\bibinfo  {journal} {Prog. Part. Nucl. Phys.}\ }\textbf {\bibinfo
  {volume} {51}},\ \bibinfo {pages} {173} (\bibinfo {year} {2003})},\ \Eprint
  {http://arxiv.org/abs/hep-ph/0212026} {arXiv:hep-ph/0212026} \BibitemShut
  {NoStop}%
\bibitem [{\citenamefont {Valcarce}\ \emph {et~al.}(1996)\citenamefont
  {Valcarce}, \citenamefont {Fernandez}, \citenamefont {Gonzalez},\ and\
  \citenamefont {Vento}}]{Valcarce:1995dm}%
  \BibitemOpen
  \bibfield  {author} {\bibinfo {author} {\bibfnamefont {A.}~\bibnamefont
  {Valcarce}}, \bibinfo {author} {\bibfnamefont {F.}~\bibnamefont {Fernandez}},
  \bibinfo {author} {\bibfnamefont {P.}~\bibnamefont {Gonzalez}}, \ and\
  \bibinfo {author} {\bibfnamefont {V.}~\bibnamefont {Vento}},\ }\href
  {\doibase 10.1016/0370-2693(95)01413-6} {\bibfield  {journal} {\bibinfo
  {journal} {Phys. Lett. B}\ }\textbf {\bibinfo {volume} {367}},\ \bibinfo
  {pages} {35} (\bibinfo {year} {1996})},\ \Eprint
  {http://arxiv.org/abs/nucl-th/9509009} {arXiv:nucl-th/9509009} \BibitemShut
  {NoStop}%
\bibitem [{\citenamefont {Vijande}\ \emph {et~al.}(2005)\citenamefont
  {Vijande}, \citenamefont {Fernandez},\ and\ \citenamefont
  {Valcarce}}]{Vijande:2004he}%
  \BibitemOpen
  \bibfield  {author} {\bibinfo {author} {\bibfnamefont {J.}~\bibnamefont
  {Vijande}}, \bibinfo {author} {\bibfnamefont {F.}~\bibnamefont {Fernandez}},
  \ and\ \bibinfo {author} {\bibfnamefont {A.}~\bibnamefont {Valcarce}},\
  }\href {\doibase 10.1088/0954-3899/31/5/017} {\bibfield  {journal} {\bibinfo
  {journal} {J. Phys. G}\ }\textbf {\bibinfo {volume} {31}},\ \bibinfo {pages}
  {481} (\bibinfo {year} {2005})},\ \Eprint
  {http://arxiv.org/abs/hep-ph/0411299} {arXiv:hep-ph/0411299} \BibitemShut
  {NoStop}%
\bibitem [{\citenamefont {Segovia}\ \emph
  {et~al.}(2008{\natexlab{a}})\citenamefont {Segovia}, \citenamefont {Entem},\
  and\ \citenamefont {Fernandez}}]{Segovia:2008zza}%
  \BibitemOpen
  \bibfield  {author} {\bibinfo {author} {\bibfnamefont {J.}~\bibnamefont
  {Segovia}}, \bibinfo {author} {\bibfnamefont {D.~R.}\ \bibnamefont {Entem}},
  \ and\ \bibinfo {author} {\bibfnamefont {F.}~\bibnamefont {Fernandez}},\
  }\href {\doibase 10.1016/j.physletb.2008.02.051} {\bibfield  {journal}
  {\bibinfo  {journal} {Phys. Lett. B}\ }\textbf {\bibinfo {volume} {662}},\
  \bibinfo {pages} {33} (\bibinfo {year} {2008}{\natexlab{a}})}\BibitemShut
  {NoStop}%
\bibitem [{\citenamefont {Segovia}\ \emph
  {et~al.}(2008{\natexlab{b}})\citenamefont {Segovia}, \citenamefont {Yasser},
  \citenamefont {Entem},\ and\ \citenamefont {Fernandez}}]{Segovia:2008zz}%
  \BibitemOpen
  \bibfield  {author} {\bibinfo {author} {\bibfnamefont {J.}~\bibnamefont
  {Segovia}}, \bibinfo {author} {\bibfnamefont {A.~M.}\ \bibnamefont {Yasser}},
  \bibinfo {author} {\bibfnamefont {D.~R.}\ \bibnamefont {Entem}}, \ and\
  \bibinfo {author} {\bibfnamefont {F.}~\bibnamefont {Fernandez}},\ }\href
  {\doibase 10.1103/PhysRevD.78.114033} {\bibfield  {journal} {\bibinfo
  {journal} {Phys. Rev. D}\ }\textbf {\bibinfo {volume} {78}},\ \bibinfo
  {pages} {114033} (\bibinfo {year} {2008}{\natexlab{b}})}\BibitemShut
  {NoStop}%
\bibitem [{\citenamefont {Segovia}\ \emph {et~al.}(2009)\citenamefont
  {Segovia}, \citenamefont {Yasser}, \citenamefont {Entem},\ and\ \citenamefont
  {Fernandez}}]{Segovia:2009zz}%
  \BibitemOpen
  \bibfield  {author} {\bibinfo {author} {\bibfnamefont {J.}~\bibnamefont
  {Segovia}}, \bibinfo {author} {\bibfnamefont {A.~M.}\ \bibnamefont {Yasser}},
  \bibinfo {author} {\bibfnamefont {D.~R.}\ \bibnamefont {Entem}}, \ and\
  \bibinfo {author} {\bibfnamefont {F.}~\bibnamefont {Fernandez}},\ }\href
  {\doibase 10.1103/PhysRevD.80.054017} {\bibfield  {journal} {\bibinfo
  {journal} {Phys. Rev. D}\ }\textbf {\bibinfo {volume} {80}},\ \bibinfo
  {pages} {054017} (\bibinfo {year} {2009})}\BibitemShut {NoStop}%
\bibitem [{\citenamefont {Segovia}\ \emph {et~al.}(2011)\citenamefont
  {Segovia}, \citenamefont {Entem},\ and\ \citenamefont
  {Fernandez}}]{Segovia:2011zza}%
  \BibitemOpen
  \bibfield  {author} {\bibinfo {author} {\bibfnamefont {J.}~\bibnamefont
  {Segovia}}, \bibinfo {author} {\bibfnamefont {D.~R.}\ \bibnamefont {Entem}},
  \ and\ \bibinfo {author} {\bibfnamefont {F.}~\bibnamefont {Fernandez}},\
  }\href {\doibase 10.1103/PhysRevD.83.114018} {\bibfield  {journal} {\bibinfo
  {journal} {Phys. Rev. D}\ }\textbf {\bibinfo {volume} {83}},\ \bibinfo
  {pages} {114018} (\bibinfo {year} {2011})}\BibitemShut {NoStop}%
\bibitem [{\citenamefont {Segovia}\ \emph {et~al.}(2015)\citenamefont
  {Segovia}, \citenamefont {Entem},\ and\ \citenamefont
  {Fernandez}}]{Segovia:2015dia}%
  \BibitemOpen
  \bibfield  {author} {\bibinfo {author} {\bibfnamefont {J.}~\bibnamefont
  {Segovia}}, \bibinfo {author} {\bibfnamefont {D.~R.}\ \bibnamefont {Entem}},
  \ and\ \bibinfo {author} {\bibfnamefont {F.}~\bibnamefont {Fernandez}},\
  }\href {\doibase 10.1103/PhysRevD.91.094020} {\bibfield  {journal} {\bibinfo
  {journal} {Phys. Rev. D}\ }\textbf {\bibinfo {volume} {91}},\ \bibinfo
  {pages} {094020} (\bibinfo {year} {2015})},\ \Eprint
  {http://arxiv.org/abs/1502.03827} {arXiv:1502.03827 [hep-ph]} \BibitemShut
  {NoStop}%
\bibitem [{\citenamefont {Ortega}\ \emph
  {et~al.}(2016{\natexlab{a}})\citenamefont {Ortega}, \citenamefont {Segovia},
  \citenamefont {Entem},\ and\ \citenamefont {Fernandez}}]{Ortega:2016mms}%
  \BibitemOpen
  \bibfield  {author} {\bibinfo {author} {\bibfnamefont {P.~G.}\ \bibnamefont
  {Ortega}}, \bibinfo {author} {\bibfnamefont {J.}~\bibnamefont {Segovia}},
  \bibinfo {author} {\bibfnamefont {D.~R.}\ \bibnamefont {Entem}}, \ and\
  \bibinfo {author} {\bibfnamefont {F.}~\bibnamefont {Fernandez}},\ }\href
  {\doibase 10.1103/PhysRevD.94.074037} {\bibfield  {journal} {\bibinfo
  {journal} {Phys. Rev. D}\ }\textbf {\bibinfo {volume} {94}},\ \bibinfo
  {pages} {074037} (\bibinfo {year} {2016}{\natexlab{a}})},\ \Eprint
  {http://arxiv.org/abs/1603.07000} {arXiv:1603.07000 [hep-ph]} \BibitemShut
  {NoStop}%
\bibitem [{\citenamefont {Segovia}\ \emph {et~al.}(2016)\citenamefont
  {Segovia}, \citenamefont {Ortega}, \citenamefont {Entem},\ and\ \citenamefont
  {Fern\'andez}}]{Segovia:2016xqb}%
  \BibitemOpen
  \bibfield  {author} {\bibinfo {author} {\bibfnamefont {J.}~\bibnamefont
  {Segovia}}, \bibinfo {author} {\bibfnamefont {P.~G.}\ \bibnamefont {Ortega}},
  \bibinfo {author} {\bibfnamefont {D.~R.}\ \bibnamefont {Entem}}, \ and\
  \bibinfo {author} {\bibfnamefont {F.}~\bibnamefont {Fern\'andez}},\ }\href
  {\doibase 10.1103/PhysRevD.93.074027} {\bibfield  {journal} {\bibinfo
  {journal} {Phys. Rev. D}\ }\textbf {\bibinfo {volume} {93}},\ \bibinfo
  {pages} {074027} (\bibinfo {year} {2016})},\ \Eprint
  {http://arxiv.org/abs/1601.05093} {arXiv:1601.05093 [hep-ph]} \BibitemShut
  {NoStop}%
\bibitem [{\citenamefont {Yang}\ \emph {et~al.}(2018)\citenamefont {Yang},
  \citenamefont {Ping},\ and\ \citenamefont {Segovia}}]{Yang:2017qan}%
  \BibitemOpen
  \bibfield  {author} {\bibinfo {author} {\bibfnamefont {G.}~\bibnamefont
  {Yang}}, \bibinfo {author} {\bibfnamefont {J.}~\bibnamefont {Ping}}, \ and\
  \bibinfo {author} {\bibfnamefont {J.}~\bibnamefont {Segovia}},\ }\href
  {\doibase 10.1007/s00601-018-1433-4} {\bibfield  {journal} {\bibinfo
  {journal} {Few Body Syst.}\ }\textbf {\bibinfo {volume} {59}},\ \bibinfo
  {pages} {113} (\bibinfo {year} {2018})},\ \Eprint
  {http://arxiv.org/abs/1709.09315} {arXiv:1709.09315 [hep-ph]} \BibitemShut
  {NoStop}%
\bibitem [{\citenamefont {Fernandez}\ \emph {et~al.}(1993)\citenamefont
  {Fernandez}, \citenamefont {Valcarce}, \citenamefont {Straub},\ and\
  \citenamefont {Faessler}}]{Fernandez:1993hx}%
  \BibitemOpen
  \bibfield  {author} {\bibinfo {author} {\bibfnamefont {F.}~\bibnamefont
  {Fernandez}}, \bibinfo {author} {\bibfnamefont {A.}~\bibnamefont {Valcarce}},
  \bibinfo {author} {\bibfnamefont {U.}~\bibnamefont {Straub}}, \ and\ \bibinfo
  {author} {\bibfnamefont {A.}~\bibnamefont {Faessler}},\ }\href {\doibase
  10.1088/0954-3899/19/12/007} {\bibfield  {journal} {\bibinfo  {journal} {J.
  Phys. G}\ }\textbf {\bibinfo {volume} {19}},\ \bibinfo {pages} {2013}
  (\bibinfo {year} {1993})}\BibitemShut {NoStop}%
\bibitem [{\citenamefont {Valcarce}\ \emph {et~al.}(1994)\citenamefont
  {Valcarce}, \citenamefont {Fernandez}, \citenamefont {Buchmann},\ and\
  \citenamefont {Faessler}}]{Valcarce:1994nr}%
  \BibitemOpen
  \bibfield  {author} {\bibinfo {author} {\bibfnamefont {A.}~\bibnamefont
  {Valcarce}}, \bibinfo {author} {\bibfnamefont {F.}~\bibnamefont {Fernandez}},
  \bibinfo {author} {\bibfnamefont {A.}~\bibnamefont {Buchmann}}, \ and\
  \bibinfo {author} {\bibfnamefont {A.}~\bibnamefont {Faessler}},\ }\href
  {\doibase 10.1103/PhysRevC.50.2246} {\bibfield  {journal} {\bibinfo
  {journal} {Phys. Rev. C}\ }\textbf {\bibinfo {volume} {50}},\ \bibinfo
  {pages} {2246} (\bibinfo {year} {1994})}\BibitemShut {NoStop}%
\bibitem [{\citenamefont {Ortega}\ \emph
  {et~al.}(2016{\natexlab{b}})\citenamefont {Ortega}, \citenamefont {Segovia},
  \citenamefont {Entem},\ and\ \citenamefont {Fern\'andez}}]{Ortega:2016hde}%
  \BibitemOpen
  \bibfield  {author} {\bibinfo {author} {\bibfnamefont {P.~G.}\ \bibnamefont
  {Ortega}}, \bibinfo {author} {\bibfnamefont {J.}~\bibnamefont {Segovia}},
  \bibinfo {author} {\bibfnamefont {D.~R.}\ \bibnamefont {Entem}}, \ and\
  \bibinfo {author} {\bibfnamefont {F.}~\bibnamefont {Fern\'andez}},\ }\href
  {\doibase 10.1103/PhysRevD.94.114018} {\bibfield  {journal} {\bibinfo
  {journal} {Phys. Rev. D}\ }\textbf {\bibinfo {volume} {94}},\ \bibinfo
  {pages} {114018} (\bibinfo {year} {2016}{\natexlab{b}})},\ \Eprint
  {http://arxiv.org/abs/1608.01325} {arXiv:1608.01325 [hep-ph]} \BibitemShut
  {NoStop}%
\bibitem [{\citenamefont {Ortega}\ \emph {et~al.}(2017)\citenamefont {Ortega},
  \citenamefont {Segovia}, \citenamefont {Entem},\ and\ \citenamefont
  {Fern\'andez}}]{Ortega:2016pgg}%
  \BibitemOpen
  \bibfield  {author} {\bibinfo {author} {\bibfnamefont {P.~G.}\ \bibnamefont
  {Ortega}}, \bibinfo {author} {\bibfnamefont {J.}~\bibnamefont {Segovia}},
  \bibinfo {author} {\bibfnamefont {D.~R.}\ \bibnamefont {Entem}}, \ and\
  \bibinfo {author} {\bibfnamefont {F.}~\bibnamefont {Fern\'andez}},\ }\href
  {\doibase 10.1103/PhysRevD.95.034010} {\bibfield  {journal} {\bibinfo
  {journal} {Phys. Rev. D}\ }\textbf {\bibinfo {volume} {95}},\ \bibinfo
  {pages} {034010} (\bibinfo {year} {2017})},\ \Eprint
  {http://arxiv.org/abs/1612.04826} {arXiv:1612.04826 [hep-ph]} \BibitemShut
  {NoStop}%
\bibitem [{\citenamefont {Ortega}\ \emph {et~al.}(2018)\citenamefont {Ortega},
  \citenamefont {Segovia}, \citenamefont {Entem},\ and\ \citenamefont
  {Fern\'andez}}]{Ortega:2017qmg}%
  \BibitemOpen
  \bibfield  {author} {\bibinfo {author} {\bibfnamefont {P.~G.}\ \bibnamefont
  {Ortega}}, \bibinfo {author} {\bibfnamefont {J.}~\bibnamefont {Segovia}},
  \bibinfo {author} {\bibfnamefont {D.~R.}\ \bibnamefont {Entem}}, \ and\
  \bibinfo {author} {\bibfnamefont {F.}~\bibnamefont {Fern\'andez}},\ }\href
  {\doibase 10.1016/j.physletb.2018.01.005} {\bibfield  {journal} {\bibinfo
  {journal} {Phys. Lett. B}\ }\textbf {\bibinfo {volume} {778}},\ \bibinfo
  {pages} {1} (\bibinfo {year} {2018})},\ \Eprint
  {http://arxiv.org/abs/1706.02639} {arXiv:1706.02639 [hep-ph]} \BibitemShut
  {NoStop}%
\bibitem [{\citenamefont {Ortega}\ \emph {et~al.}(2019)\citenamefont {Ortega},
  \citenamefont {Segovia}, \citenamefont {Entem},\ and\ \citenamefont
  {Fern\'andez}}]{Ortega:2018cnm}%
  \BibitemOpen
  \bibfield  {author} {\bibinfo {author} {\bibfnamefont {P.~G.}\ \bibnamefont
  {Ortega}}, \bibinfo {author} {\bibfnamefont {J.}~\bibnamefont {Segovia}},
  \bibinfo {author} {\bibfnamefont {D.~R.}\ \bibnamefont {Entem}}, \ and\
  \bibinfo {author} {\bibfnamefont {F.}~\bibnamefont {Fern\'andez}},\ }\href
  {\doibase 10.1140/epjc/s10052-019-6552-7} {\bibfield  {journal} {\bibinfo
  {journal} {Eur. Phys. J. C}\ }\textbf {\bibinfo {volume} {79}},\ \bibinfo
  {pages} {78} (\bibinfo {year} {2019})},\ \Eprint
  {http://arxiv.org/abs/1808.00914} {arXiv:1808.00914 [hep-ph]} \BibitemShut
  {NoStop}%
\bibitem [{\citenamefont {Ortega}\ \emph {et~al.}(2020)\citenamefont {Ortega},
  \citenamefont {Segovia}, \citenamefont {Entem},\ and\ \citenamefont
  {Fernandez}}]{Ortega:2020uvc}%
  \BibitemOpen
  \bibfield  {author} {\bibinfo {author} {\bibfnamefont {P.~G.}\ \bibnamefont
  {Ortega}}, \bibinfo {author} {\bibfnamefont {J.}~\bibnamefont {Segovia}},
  \bibinfo {author} {\bibfnamefont {D.~R.}\ \bibnamefont {Entem}}, \ and\
  \bibinfo {author} {\bibfnamefont {F.}~\bibnamefont {Fernandez}},\ }\href
  {\doibase 10.1140/epjc/s10052-020-7764-6} {\bibfield  {journal} {\bibinfo
  {journal} {Eur. Phys. J. C}\ }\textbf {\bibinfo {volume} {80}},\ \bibinfo
  {pages} {223} (\bibinfo {year} {2020})},\ \Eprint
  {http://arxiv.org/abs/2001.08093} {arXiv:2001.08093 [hep-ph]} \BibitemShut
  {NoStop}%
\bibitem [{\citenamefont {Vijande}\ \emph {et~al.}(2006)\citenamefont
  {Vijande}, \citenamefont {Valcarce},\ and\ \citenamefont
  {Tsushima}}]{Vijande:2006jf}%
  \BibitemOpen
  \bibfield  {author} {\bibinfo {author} {\bibfnamefont {J.}~\bibnamefont
  {Vijande}}, \bibinfo {author} {\bibfnamefont {A.}~\bibnamefont {Valcarce}}, \
  and\ \bibinfo {author} {\bibfnamefont {K.}~\bibnamefont {Tsushima}},\ }\href
  {\doibase 10.1103/PhysRevD.74.054018} {\bibfield  {journal} {\bibinfo
  {journal} {Phys. Rev. D}\ }\textbf {\bibinfo {volume} {74}},\ \bibinfo
  {pages} {054018} (\bibinfo {year} {2006})},\ \Eprint
  {http://arxiv.org/abs/hep-ph/0608316} {arXiv:hep-ph/0608316} \BibitemShut
  {NoStop}%
\bibitem [{\citenamefont {Yang}\ and\ \citenamefont
  {Ping}(2017)}]{Yang:2015bmv}%
  \BibitemOpen
  \bibfield  {author} {\bibinfo {author} {\bibfnamefont {G.}~\bibnamefont
  {Yang}}\ and\ \bibinfo {author} {\bibfnamefont {J.}~\bibnamefont {Ping}},\
  }\href {\doibase 10.1103/PhysRevD.95.014010} {\bibfield  {journal} {\bibinfo
  {journal} {Phys. Rev. D}\ }\textbf {\bibinfo {volume} {95}},\ \bibinfo
  {pages} {014010} (\bibinfo {year} {2017})},\ \Eprint
  {http://arxiv.org/abs/1511.09053} {arXiv:1511.09053 [hep-ph]} \BibitemShut
  {NoStop}%
\bibitem [{\citenamefont {Yang}\ \emph {et~al.}(2019)\citenamefont {Yang},
  \citenamefont {Ping},\ and\ \citenamefont {Segovia}}]{Yang:2018oqd}%
  \BibitemOpen
  \bibfield  {author} {\bibinfo {author} {\bibfnamefont {G.}~\bibnamefont
  {Yang}}, \bibinfo {author} {\bibfnamefont {J.}~\bibnamefont {Ping}}, \ and\
  \bibinfo {author} {\bibfnamefont {J.}~\bibnamefont {Segovia}},\ }\href
  {\doibase 10.1103/PhysRevD.99.014035} {\bibfield  {journal} {\bibinfo
  {journal} {Phys. Rev. D}\ }\textbf {\bibinfo {volume} {99}},\ \bibinfo
  {pages} {014035} (\bibinfo {year} {2019})},\ \Eprint
  {http://arxiv.org/abs/1809.06193} {arXiv:1809.06193 [hep-ph]} \BibitemShut
  {NoStop}%
\bibitem [{\citenamefont {Yang}\ \emph
  {et~al.}(2020{\natexlab{a}})\citenamefont {Yang}, \citenamefont {Ping},\ and\
  \citenamefont {Segovia}}]{Yang:2019itm}%
  \BibitemOpen
  \bibfield  {author} {\bibinfo {author} {\bibfnamefont {G.}~\bibnamefont
  {Yang}}, \bibinfo {author} {\bibfnamefont {J.}~\bibnamefont {Ping}}, \ and\
  \bibinfo {author} {\bibfnamefont {J.}~\bibnamefont {Segovia}},\ }\href
  {\doibase 10.1103/PhysRevD.101.014001} {\bibfield  {journal} {\bibinfo
  {journal} {Phys. Rev. D}\ }\textbf {\bibinfo {volume} {101}},\ \bibinfo
  {pages} {014001} (\bibinfo {year} {2020}{\natexlab{a}})},\ \Eprint
  {http://arxiv.org/abs/1911.00215} {arXiv:1911.00215 [hep-ph]} \BibitemShut
  {NoStop}%
\bibitem [{\citenamefont {Yang}\ \emph
  {et~al.}(2020{\natexlab{b}})\citenamefont {Yang}, \citenamefont {Ping},\ and\
  \citenamefont {Segovia}}]{Yang:2020twg}%
  \BibitemOpen
  \bibfield  {author} {\bibinfo {author} {\bibfnamefont {G.}~\bibnamefont
  {Yang}}, \bibinfo {author} {\bibfnamefont {J.}~\bibnamefont {Ping}}, \ and\
  \bibinfo {author} {\bibfnamefont {J.}~\bibnamefont {Segovia}},\ }\href
  {\doibase 10.1103/PhysRevD.101.074030} {\bibfield  {journal} {\bibinfo
  {journal} {Phys. Rev. D}\ }\textbf {\bibinfo {volume} {101}},\ \bibinfo
  {pages} {074030} (\bibinfo {year} {2020}{\natexlab{b}})},\ \Eprint
  {http://arxiv.org/abs/2003.05253} {arXiv:2003.05253 [hep-ph]} \BibitemShut
  {NoStop}%
\bibitem [{\citenamefont {Yang}\ \emph
  {et~al.}(2020{\natexlab{c}})\citenamefont {Yang}, \citenamefont {Ping},\ and\
  \citenamefont {Segovia}}]{Yang:2020fou}%
  \BibitemOpen
  \bibfield  {author} {\bibinfo {author} {\bibfnamefont {G.}~\bibnamefont
  {Yang}}, \bibinfo {author} {\bibfnamefont {J.}~\bibnamefont {Ping}}, \ and\
  \bibinfo {author} {\bibfnamefont {J.}~\bibnamefont {Segovia}},\ }\href
  {\doibase 10.1103/PhysRevD.102.054023} {\bibfield  {journal} {\bibinfo
  {journal} {Phys. Rev. D}\ }\textbf {\bibinfo {volume} {102}},\ \bibinfo
  {pages} {054023} (\bibinfo {year} {2020}{\natexlab{c}})},\ \Eprint
  {http://arxiv.org/abs/2007.05190} {arXiv:2007.05190 [hep-ph]} \BibitemShut
  {NoStop}%
\bibitem [{\citenamefont {Yang}\ \emph
  {et~al.}(2020{\natexlab{d}})\citenamefont {Yang}, \citenamefont {Ping},\ and\
  \citenamefont {Segovia}}]{Yang:2020atz}%
  \BibitemOpen
  \bibfield  {author} {\bibinfo {author} {\bibfnamefont {G.}~\bibnamefont
  {Yang}}, \bibinfo {author} {\bibfnamefont {J.}~\bibnamefont {Ping}}, \ and\
  \bibinfo {author} {\bibfnamefont {J.}~\bibnamefont {Segovia}},\ }\href
  {\doibase 10.3390/sym12111869} {\bibfield  {journal} {\bibinfo  {journal}
  {Symmetry}\ }\textbf {\bibinfo {volume} {12}},\ \bibinfo {pages} {1869}
  (\bibinfo {year} {2020}{\natexlab{d}})},\ \Eprint
  {http://arxiv.org/abs/2009.00238} {arXiv:2009.00238 [hep-ph]} \BibitemShut
  {NoStop}%
\bibitem [{\citenamefont {Yang}\ \emph {et~al.}(2021)\citenamefont {Yang},
  \citenamefont {Ping},\ and\ \citenamefont {Segovia}}]{Yang:2021izl}%
  \BibitemOpen
  \bibfield  {author} {\bibinfo {author} {\bibfnamefont {G.}~\bibnamefont
  {Yang}}, \bibinfo {author} {\bibfnamefont {J.}~\bibnamefont {Ping}}, \ and\
  \bibinfo {author} {\bibfnamefont {J.}~\bibnamefont {Segovia}},\ }\href
  {\doibase 10.1103/PhysRevD.103.074011} {\bibfield  {journal} {\bibinfo
  {journal} {Phys. Rev. D}\ }\textbf {\bibinfo {volume} {103}},\ \bibinfo
  {pages} {074011} (\bibinfo {year} {2021})},\ \Eprint
  {http://arxiv.org/abs/2101.04933} {arXiv:2101.04933 [hep-ph]} \BibitemShut
  {NoStop}%
\bibitem [{\citenamefont {Hammond}\ \emph {et~al.}(1994)\citenamefont
  {Hammond}, \citenamefont {Lester},\ and\ \citenamefont
  {Reynolds}}]{Hammond:1994bk}%
  \BibitemOpen
  \bibfield  {author} {\bibinfo {author} {\bibfnamefont {B.}~\bibnamefont
  {Hammond}}, \bibinfo {author} {\bibfnamefont {W.}~\bibnamefont {Lester}}, \
  and\ \bibinfo {author} {\bibfnamefont {P.}~\bibnamefont {Reynolds}},\
  }\href@noop {} {\emph {\bibinfo {title} {Monte Carlo Methods in ab Initio
  Quantum Chemistry}}}\ (\bibinfo  {publisher} {World Scientific},\ \bibinfo
  {address} {Singapore},\ \bibinfo {year} {1994})\BibitemShut {NoStop}%
\bibitem [{\citenamefont {Foulkes}\ \emph {et~al.}(2001)\citenamefont
  {Foulkes}, \citenamefont {Mitas}, \citenamefont {Needs},\ and\ \citenamefont
  {Rajagopal}}]{Foulkes:2001zz}%
  \BibitemOpen
  \bibfield  {author} {\bibinfo {author} {\bibfnamefont {W.}~\bibnamefont
  {Foulkes}}, \bibinfo {author} {\bibfnamefont {L.}~\bibnamefont {Mitas}},
  \bibinfo {author} {\bibfnamefont {R.}~\bibnamefont {Needs}}, \ and\ \bibinfo
  {author} {\bibfnamefont {G.}~\bibnamefont {Rajagopal}},\ }\href {\doibase
  10.1103/RevModPhys.73.33} {\bibfield  {journal} {\bibinfo  {journal} {Rev.
  Mod. Phys.}\ }\textbf {\bibinfo {volume} {73}},\ \bibinfo {pages} {33}
  (\bibinfo {year} {2001})}\BibitemShut {NoStop}%
\bibitem [{\citenamefont {Nightingale}\ and\ \citenamefont
  {Umrigar}(2014)}]{Nightingale:2014bk}%
  \BibitemOpen
  \bibfield  {author} {\bibinfo {author} {\bibfnamefont {M.}~\bibnamefont
  {Nightingale}}\ and\ \bibinfo {author} {\bibfnamefont {C.~J.}\ \bibnamefont
  {Umrigar}},\ }\href@noop {} {\emph {\bibinfo {title} {Quantum Monte Carlo
  Methods in Physics and Chemistry}}}\ (\bibinfo  {publisher} {Springer},\
  \bibinfo {address} {Vienna},\ \bibinfo {year} {2014})\BibitemShut {NoStop}%
\bibitem [{\citenamefont {Carlson}\ \emph
  {et~al.}(1983{\natexlab{a}})\citenamefont {Carlson}, \citenamefont {Kogut},\
  and\ \citenamefont {Pandharipande}}]{Carlson:1982xi}%
  \BibitemOpen
  \bibfield  {author} {\bibinfo {author} {\bibfnamefont {J.}~\bibnamefont
  {Carlson}}, \bibinfo {author} {\bibfnamefont {J.~B.}\ \bibnamefont {Kogut}},
  \ and\ \bibinfo {author} {\bibfnamefont {V.}~\bibnamefont {Pandharipande}},\
  }\href {\doibase 10.1103/PhysRevD.27.233} {\bibfield  {journal} {\bibinfo
  {journal} {Phys. Rev. D}\ }\textbf {\bibinfo {volume} {27}},\ \bibinfo
  {pages} {233} (\bibinfo {year} {1983}{\natexlab{a}})}\BibitemShut {NoStop}%
\bibitem [{\citenamefont {Carlson}\ \emph
  {et~al.}(1983{\natexlab{b}})\citenamefont {Carlson}, \citenamefont {Kogut},\
  and\ \citenamefont {Pandharipande}}]{Carlson:1983rw}%
  \BibitemOpen
  \bibfield  {author} {\bibinfo {author} {\bibfnamefont {J.}~\bibnamefont
  {Carlson}}, \bibinfo {author} {\bibfnamefont {J.}~\bibnamefont {Kogut}}, \
  and\ \bibinfo {author} {\bibfnamefont {V.}~\bibnamefont {Pandharipande}},\
  }\href {\doibase 10.1103/PhysRevD.28.2807} {\bibfield  {journal} {\bibinfo
  {journal} {Phys. Rev. D}\ }\textbf {\bibinfo {volume} {28}},\ \bibinfo
  {pages} {2807} (\bibinfo {year} {1983}{\natexlab{b}})}\BibitemShut {NoStop}%
\bibitem [{\citenamefont {Isgur}\ and\ \citenamefont
  {Karl}(1978)}]{Isgur:1978xj}%
  \BibitemOpen
  \bibfield  {author} {\bibinfo {author} {\bibfnamefont {N.}~\bibnamefont
  {Isgur}}\ and\ \bibinfo {author} {\bibfnamefont {G.}~\bibnamefont {Karl}},\
  }\href {\doibase 10.1103/PhysRevD.18.4187} {\bibfield  {journal} {\bibinfo
  {journal} {Phys. Rev. D}\ }\textbf {\bibinfo {volume} {18}},\ \bibinfo
  {pages} {4187} (\bibinfo {year} {1978})}\BibitemShut {NoStop}%
\bibitem [{\citenamefont {Isgur}\ and\ \citenamefont
  {Karl}(1979{\natexlab{a}})}]{Isgur:1978wd}%
  \BibitemOpen
  \bibfield  {author} {\bibinfo {author} {\bibfnamefont {N.}~\bibnamefont
  {Isgur}}\ and\ \bibinfo {author} {\bibfnamefont {G.}~\bibnamefont {Karl}},\
  }\href {\doibase 10.1103/PhysRevD.19.2653} {\bibfield  {journal} {\bibinfo
  {journal} {Phys. Rev. D}\ }\textbf {\bibinfo {volume} {19}},\ \bibinfo
  {pages} {2653} (\bibinfo {year} {1979}{\natexlab{a}})},\ \bibinfo {note}
  {[Erratum: Phys.Rev.D 23, 817 (1981)]}\BibitemShut {NoStop}%
\bibitem [{\citenamefont {Isgur}\ and\ \citenamefont
  {Karl}(1979{\natexlab{b}})}]{Isgur:1979be}%
  \BibitemOpen
  \bibfield  {author} {\bibinfo {author} {\bibfnamefont {N.}~\bibnamefont
  {Isgur}}\ and\ \bibinfo {author} {\bibfnamefont {G.}~\bibnamefont {Karl}},\
  }\href {\doibase 10.1103/PhysRevD.20.1191} {\bibfield  {journal} {\bibinfo
  {journal} {Phys. Rev. D}\ }\textbf {\bibinfo {volume} {20}},\ \bibinfo
  {pages} {1191} (\bibinfo {year} {1979}{\natexlab{b}})}\BibitemShut {NoStop}%
\bibitem [{\citenamefont {Capstick}\ and\ \citenamefont
  {Isgur}(1985)}]{Capstick:1986bm}%
  \BibitemOpen
  \bibfield  {author} {\bibinfo {author} {\bibfnamefont {S.}~\bibnamefont
  {Capstick}}\ and\ \bibinfo {author} {\bibfnamefont {N.}~\bibnamefont
  {Isgur}},\ }\href {\doibase 10.1103/PhysRevD.34.2809} {\bibfield  {journal}
  {\bibinfo  {journal} {AIP Conf. Proc.}\ }\textbf {\bibinfo {volume} {132}},\
  \bibinfo {pages} {267} (\bibinfo {year} {1985})}\BibitemShut {NoStop}%
\bibitem [{\citenamefont {Gordillo}\ \emph {et~al.}(2020)\citenamefont
  {Gordillo}, \citenamefont {De~Soto},\ and\ \citenamefont
  {Segovia}}]{Gordillo:2020sgc}%
  \BibitemOpen
  \bibfield  {author} {\bibinfo {author} {\bibfnamefont {M.~C.}\ \bibnamefont
  {Gordillo}}, \bibinfo {author} {\bibfnamefont {F.}~\bibnamefont {De~Soto}}, \
  and\ \bibinfo {author} {\bibfnamefont {J.}~\bibnamefont {Segovia}},\ }\href
  {\doibase 10.1103/PhysRevD.102.114007} {\bibfield  {journal} {\bibinfo
  {journal} {Phys. Rev. D}\ }\textbf {\bibinfo {volume} {102}},\ \bibinfo
  {pages} {114007} (\bibinfo {year} {2020})},\ \Eprint
  {http://arxiv.org/abs/2009.11889} {arXiv:2009.11889 [hep-ph]} \BibitemShut
  {NoStop}%
\end{thebibliography}%

\end{document}